 \documentclass[twocolumn,tighten]{aastex62}
\newcommand{\atl}{ATLAS$^{\rm 3D}$}
\newcommand{\kms}{km~s$^{-1}$}
\newcommand{\siglos}{$\sigma_{\rm los}(R)$}

\usepackage{amsmath}

\shorttitle{Orbital Anisotropy of Slow Rotators}
\shortauthors{Chae et al.}

\begin{document}

\title{Modeling Nearly Spherical Pure-Bulge Galaxies with a Stellar Mass-to-Light Ratio Gradient under the $\Lambda$CDM and MOND Paradigms: II. The Orbital Anisotropy of Slow Rotators within the Effective Radius}

\correspondingauthor{Kyu-Hyun Chae}
\email{chae@sejong.ac.kr, kyuhyunchae@gmail.com}

\author{Kyu-Hyun Chae}
\affiliation{Department of Physics and Astronomy, Sejong University, 
 209 Neungdong-ro Gwangjin-gu, Seoul 05006, Republic of Korea}

\author{Mariangela Bernardi}
\affil{Department of Physics and Astronomy, University of Pennsylvania,
209 South 33rd Street, Philadelphia, PA 19104, USA}

\author{Ravi K. Sheth}
\affil{Department of Physics and Astronomy, University of Pennsylvania,
209 South 33rd Street, Philadelphia, PA 19104, USA}



\begin{abstract}

  We investigate the anisotropy of the stellar velocity dispersions within the effective radius, $R_{\rm e}$, in 24 {\atl} pure-bulge galaxies, 16 of which are kinematic slow rotators (SRs). We allow the spherical anisotropy parameter $\beta$ to be radially varying and allow a radial gradient in the stellar mass-to-light ratio ($M_\star/L$) through the parameter $K$ introduced earlier. The median anisotropy for SRs depends on $K$ as follows: $\langle\beta_{\rm m}\rangle = a + b K$ with $a=0.19\pm 0.05$, $b=-0.13\pm 0.07$ ($\Lambda$CDM) or $a=0.21\pm 0.05$, $b=-0.26\pm 0.08$ (MOND), where $\beta_{\rm m}$ refers to the radially averaged quantity. Under the $\Lambda$CDM paradigm this scaling is tied to a scaling of $\langle f_{\rm DM}\rangle = (0.16\pm 0.03) +(0.31\pm 0.06) K$, where $f_{\rm DM}$ refers to the DM fraction within a sphere of $r=R_{\rm e}$. For $K=0$ (constant $M_\star/L$), we obtain radially biased results with $\langle\beta_{\rm m}\rangle \approx 0.2$ consistent with previous results. However, marginalizing over $0 < K < 1.5$ yields $\langle\beta_{\rm m}\rangle = 0.06 ^{+0.11}_{-0.14}$ with $\langle f_{\rm DM}\rangle = 0.35 \pm 0.08$: isotropy is preferred. This isotropy hides the fact that $\beta_{\rm m}$ is correlated with kinematic features such as counter rotating cores (CRCs), kinematically distinct cores (KDCs), and low-level velocities (LVs): SRs with LVs are likely to be radially biased while SRs with CRCs are likely to be tangentially biased, and SRs with KDCs are intermediate. Existing cosmological simulations allow us to understand these results qualitatively in terms of their dynamical structures and formation histories although there exist quantitative tensions. More realistic cosmological simulations, particularly allowing for $M_\star/L$ gradients, may be required to better understand SRs.

\end{abstract}




\section{Introduction} \label{sec:intro}

Observed galaxies exhibit a great variety in appearances, constituents and kinematic properties of stars (and gas particles). In galaxies, gravitationally bound stellar orbits can be realized in a number of possible ways including (rotational) circular orbits, radial orbits, box orbits, tube orbits, irregular/chaotic orbits, etc (see, e.g., \citealt{deZ85,Stat87,BT,Rott14}). Based on the overall properties of the orbits, galaxies are broadly referred to as being rotationally supported (or dominated) if circular orbits dominate as in disk galaxies, or dispersion/pressure supported (or dominated) otherwise. Furthermore, a galaxy, whether it is rotation or dispersion dominated, can contain several kinematically distinct sub-systems including a rotating disk, a dispersion supported bulge, and a dispersion supported dark matter halo among others (see, e.g., \citealt{BT,Cap16,Kor16} and references therein). 

The use of integral field spectroscopy (IFS) for kinematic studies of galaxies for the past two decades, led by the SAURON \citep{deZ02} and {\atl} \citep{Cap11} surveys, has revealed crucial aspects and details of the structure and dynamics of early-type (i.e.\ lenticular and elliptical) galaxies (ETGs) \citep{Ems07,Cap07,Kra11,Ems11,Cap13}. Strikingly, the vast majority of not only lenticular but also elliptical galaxies exhibit varying degrees of rotation \citep{Ems07,Ems11}, and also contain disks as revealed by photometric decomposition \citep{Kra13}. When classified by the angular momentum parameter $\lambda_R$ introduced by \citet{Ems07}, only $14\pm 2$ \% of the {\atl} sample (36 out of 260) have $\lambda_{\rm e} < 0.31\sqrt{\varepsilon_{\rm e}}$ (where $\varepsilon_{\rm e}$ is the ellipticity of the observed light distribution) showing little or no rotation within the effective radius $R_{\rm e}$ (the half light radius in the projected light distribution). This minority is referred to collectively as slow rotators (SRs), while all the rest are fast rotators (FRs) \citep{Ems11}. 

SRs appear nearly round, relatively more massive, tend to have irregular kinematics \citep{Kra08,Kra11}, and do not usually possess detectable disks \citep{Ems11,Kra13}. The last property means that what appear to be pure-bulges on the basis of photometry are likely to be kinematic SRs. In our selection \citep{CBS18a} (hereafter Paper~I) two thirds of pure-bulges (16 out of 24) are SRs. We have carried out spherical Jeans modeling of 24 {\atl} pure-bulge galaxies to address multiple astrophysical issues including the radial acceleration relation (RAR) in a super-critical acceleration regime from $\sim 10^{-9.5}~{\rm m}~{\rm s}^{-2}$ - $\sim 10^{-8}~{\rm m}~{\rm s}^{-2}$ \citep{CBS18b}, galactic structure, and the distribution of stellar orbits. The last point is the subject of this paper. 

All 260 {\atl} galaxies have been modeled and analyzed through the Jeans Anisotropic Modeling (JAM) code by the {\atl} team \citep{Cap13}. Our modeling (Paper I) is different from the JAM analysis in several ways. First of all, we allow a radial gradient in the stellar mass-to-light ratio ($M_\star/L$); this gradient is confined to the central region ($<0.4 R_{\rm e}$), with a parameter $K$ representing the strength of a possible gradient.  This is motivated by multiple recent reports that the Initial Mass Function (IMF) of stars in the central regions of ETGs is bottom heavy (e.g.\ \citealt{MN15,LaB16,vD17,Sar18,Son18}) and this drives a gradient in $M_\star/L$.  It should be noted, however, that there are ETGs in which IMF gradients have not been found (e.g.\ \citealt{Alt17,Alt18,DM17}). Systematic errors in interpreting the relevant spectral features may be responsible for these non-convergent results, but they may also reflect intrinsic galaxy-to-galaxy variations. Interestingly, Paper~I finds that posterior inferences of $K$ for the 24 pure-bulges exhibit large scatter, with the median strength $\langle K \rangle \sim 0.55$ lying between $K=1$ and $K=0$, which represent, respectively, the strong gradient reported by \citet{vD17}, and the case of no gradient at all reported by \citet{Alt18}. It is also interesting to note that the Paper~I inference of the gradient for NGC~4486 (M87) agrees well with recent independent studies of the galaxy by \citet{Old18} and \citet{Sar18}.

Secondly, we allow for radially varying spherical anisotropies $\beta(r)$ in the stellar velocity dispersions, by using a generalized Osipkov-Merritt \citep{Osip,Merr} model for the region probed by IFS data (typically $r \la R_{\rm e}$). This was an effort to improve model-fitting of the IFS data, but also to investigate possible radial variation that can be compared with theoretical (cosmological hydrodynamic simulation) predictions.

Thirdly, in the $\Lambda$CDM paradigm (e.g., \citealt{MBW10}), the stellar dynamics may be affected by the presence of dark matter (DM).  Whereas the standard lore has been that DM matters little on the small scales currently probed by IFS studies, \cite{Ber18} made the point that if $M_\star/L$ increases towards the center, then this tends to increase the required contribution from DM.  Therefore, because we are considering $M_\star/L$ gradients, we allow for different parameterizations of the DM distribution associated with the halo of a galaxy.  In what follows, we use DM profiles motivated by \cite{Ein} and \cite{NFW}.

Fourth, as there is ongoing discussion (see, e.g., \citet{Jan16} for a study of some \atl\ FRs assuming constant $M_\star/L$ and constant anisotropy) of whether effects attributed to a DM halo can instead be explained by modified Newtonian dynamics (MOND: \citealt{Mil}), it is interesting to also consider the MOND paradigm for our study of orbital anisotropies in the presence of $M_*/L$ gradients.  This is despite the fact that, in these galaxies, the radial acceleration due to baryons ranges from $\sim 10^{-9.5}$~m~s$^{-2}$ to $\sim 10^{-8}$~m~s$^{-2}$ \citep{CBS18b}, which is considerably larger than the critical acceleration, $a_0 \sim 10^{-10}$~m~s$^{-2}$, at which MOND is usually invoked.  However, MONDian effects depend on the precise shape of the MOND interpolating function (IF) (see \citealt{CBS18b}), and for some IFs, MONDian effects can be non-negligible, especially if $M_\star/L$ gradients in our sample are significant. Hence we consider two different families of MONDian interpolating functions.  

Therefore, we can here constrain the velocity dispersion anisotropies of pure-bulge SRs -- which we will refer to as nearly Spherical, Slowly-rotating, pure-Bulge Galaxies (SSBGs) --  in an unprecedented way, taking account of the effects of $M_\star/L$ radial gradients. The velocity dispersion anisotropy is a key parameter characterizing the dynamics of SSBGs and can provide useful constraints on theories of galaxy formation and evolution. As shown in recent state-of-the-art cosmological hydrodynamic simulations (e.g.\ \citealt{Naab14,Rott14,Wu14,Xu17,Li18}) as well as galaxy merger simulations (e.g.\ \citealt{BQ90,BH96,Jess05,Jess07,Bois11,Hilz12,Tsat15}) under the $\Lambda$CDM paradigm, the assembly history of a galaxy is closely linked with the composition of its stellar orbits, the global angular momentum (i.e.\ whether it is a fast or slow rotator), and the velocity dispersion anisotropy profile as well as various morphological and photometric properties. A concise and excellent account of the current state of theoretical ideas and simulations can be found, e.g., in \S 2 of \citet{Naab14}. SSBGs may look simple but their present states may be the outcome of complex histories involving in-situ star formation, dry (or gas-rich) major or minor mergers, continual accretion and feedback from supernovae and AGNs. Empirically determined velocity dispersion anisotropies may constrain such processes. Simulations (e.g.\ \citealt{Rott14,Wu14,Xu17}) suggest, in particular, that the fraction of in-situ stars is well-correlated with the anisotropy, in the sense that when the in-situ fraction is smaller, the orbital distribution is more dominated by radial orbits (i.e., accreted stars tend to be radially biased).

This paper is structured as follows. In \S~\ref{sec:model}, we briefly describe the spherical Jeans Monte Carlo model ingredients that are directly relevant to this work, while referring to Paper~I (and also \citet{CBS18b}) for details. We present our estimates of the velocity dispersion anisotropies of 16 SSBGs drawn from the {\atl} sample in \S~\ref{sec:result}, and show that the $M_\star/L$ radial gradient is a non-negligible factor in inferring anisotropies from dynamical analyses of SSBGs. In \S~\ref{sec:sim}, we compare our results on the velocity dispersion anisotropy with the predictions by currently available cosmological simulations. We discuss our results and conclude in \S~\ref{sec:conc}. In the Appendix, we provide fitted values of the anisotropy, $M_\star/L$ and $f_{\rm DM}$ (the DM fraction within $R_{\rm e}$).

\section{Model Ingredients} \label{sec:model}

In our approach of using the spherical Jeans equation (\citealt{BT}, Equation~(4.215)) we do not construct a library of orbits.  Rather, we use the anisotropy parameter 
\begin{equation}
\beta = 1 - \frac{\bar{v_\theta^2}+\bar{v_\phi^2}}{2 \bar{v_{\rm r}^2}},
\label{eq:bet} 
\end{equation}
to gain information about the distribution of orbits at a given point. Here $\bar{v_{\rm r}^2}$, $\bar{v_\theta^2}$, and $\bar{v_\phi^2}$ are the mean squared velocities (``second moments'') in the spherical coordinates. These are equivalent to velocity dispersions $\sigma_{\rm r}^2$, etc. for non-rotating systems, e.g.\ the SSBGs considered here. If $\beta > 0$ ($\beta < 0$), the velocity dispersions are radially (tangentially) biased. We allow a radial variation of $\beta$ even for the relatively small regions ($\la R_{\rm e}$) probed by the IFS observations in the optical. For this, we use the generalized Osipkov-Merritt (gOM) model (\citealt{Osip,Merr}; see also \citet{BT}, p.\ 297), 
\begin{equation}
  \beta_{\rm gOM}(r)=\beta_0 + (\beta_\infty - \beta_0) \frac{(r/r_a)^2}{1+(r/r_a)^2},
 \label{eq:gOM}
\end{equation}
which varies smoothly from a central value $\beta_0$ to $\beta_\infty$ at large radii. We consider the range $[-2, 0.7]$ for both $\beta_0$ and $\beta_\infty$ so that $1/3 \le \sigma_{\rm t}^2/\sigma_{\rm r}^2 \le 3$ where $\sigma_{\rm t}^2 = (\sigma_{\theta}^2+\sigma_{\phi}^2)/2$ is the one-dimensional tangential velocity dispersion. For the scale parameter $r_a$ we consider the range $0<r_a<R_{\rm e}$ relevant to the probed region. The model given by Equation~(\ref{eq:gOM}) is intended to approximate smoothly realistic anisotropy profiles which could be obtained from orbit superposition methods (e.g., \citealt{RT88,vdM98,Ger01,Geb03,Kra05,Tho07,Cap07}).

The stellar mass distribution in a galaxy is obtained by multiplying the observed light distribution by a stellar mass-to-light ratio $\Upsilon_\star \equiv M_\star/L$. We parameterize this by
\begin{equation}
\frac{\Upsilon_\star (R/R_{\rm e})}{\Upsilon_{\star 0}} =\max\left\{1+ K\left[A - B (R/R_{\rm e})\right],1\right\},
 \label{eq:MLgrad}
\end{equation}
where $R$ is a separation projected along the line of sight onto the plane of the sky, and
$(A,~B)=(2.33,~6.00)$ are derived by \cite{Ber18} for the recently observed $M_\star/L$ gradient \citep{vD17}. Current observational results (e.g., \citealt{MN15,LaB16,vD17,Sar18,Son18,Alt18}) correspond to the range $0\la K \la 1$. In what follows, we consider two separate analyses: one in which there is no gradient ($K=0$) and another in which $K$ is allowed to be in the range $[0,1.5]$.

The JAM modeling results of the \atl\ ETGs, based on the assumption of constant $M_\star/L$ ($K=0$ in our language) show that the inner regions within $R_{\rm e}$ are dominated by baryons with a median DM fraction of $\sim 13$\% within a spherical radius of $R_{\rm e}$ \citep{Cap13}. However, as pointed out by \citet{Ber18}, if $M_\star/L$ increases towards the center ($K>0$ in our language) due to a radial variation in stellar populations or IMF, then one expects larger DM fractions to be required. This is because the change in the baryon distribution (by the radial variation of $M_\star/L$) necessarily requires an adjustment in the DM distribution to obtain the correct total mass distribution for the observed stellar dynamics (velocity dispersions here). That is to say, analyses that assume $K=0$ underestimate the importance of a DM halo in the observed stellar dynamics within $R_{\rm e}$.

\subsection{Dark matter models}
We consider two classes of models that can describe the smooth distribution of DM. These are generalizations of the DM-only simulation prediction (e.g., \citealt{NFW,Merr06,Nav10}) that would allow for modifications caused by baryonic and feedback physics. One is a generalized NFW (gNFW) density profile,
\begin{equation}
\rho_{\rm gNFW}(r)=\frac{\rho_{\rm s}} {\left(r/r_{\rm s}\right)^\alpha 
\left(1+r/r_{\rm s}\right)^{3-\alpha}},
 \label{eq:gNFW} 
\end{equation}
where $-\alpha$ is the inner density power-law slope ($\alpha=1$ being the NFW case) and $r_{\rm s}$ is the scale radius. We consider the range $0.1 < \alpha < 1.8$. The other is the Einasto profile, 
\begin{equation}
\rho_{\rm Ein}(r)= \rho_{-2} 
\exp\left\{-(2/\tilde{\alpha})\left[(r/r_{-2})^{\tilde{\alpha}}-1\right]\right\},
 \label{eq:Ein} 
\end{equation}
where $r_{-2}$ is the radius at which the logarithmic slope of the density is $-2$ and controls the slope variation with radius along with $\tilde{\alpha}$ as follows: $\gamma_{\rm Ein}\equiv d\ln\rho_{\rm Ein}(r)/d\ln r = -2(r/r_{-2})^{\tilde{\alpha}}$. N-body simulations give $0.15 \la \tilde{\alpha} \la 0.2$ (e.g., \citealt{Nav10,Merr06}). Matching the NFW profile with the Einasto profile with $\tilde{\alpha}=0.17$ we obtain the radius where the slope is $-1$ as $r_{-1}/r_{-2}=0.017$. Now, the slope at this fiducial radius $r_{-1}=0.017 r_{-2}$ is related to $\tilde{\alpha}$ as $\tilde{\alpha}=\ln(-\gamma_{\rm Ein}/2)/\ln (0.017)$. In obtaining a Monte Carlo (MC) set of models we take a uniform deviate of $\gamma_{\rm Ein}$ from $(0.1,1.8)$, which corresponds to a range $0.025 < \tilde{\alpha} < 0.74$, rather than taking $\tilde{\alpha}$ directly from the range as it will be biased against smaller values of $\tilde{\alpha}$.

\subsection{MONDian models}
Models in the MOND paradigm are distinguished by the relation $a/a_{\rm B}=f(a_{\rm B}/a_0)$ where $a$ is the actual acceleration, $a_{\rm B}$ is the acceleration predicted by the distribution of baryons (stars here) based on Newtonian dynamics, and $a_0\approx 1.2 \times 10^{-10}$ m~s$^{-2}$ is the critical acceleration.
The function $f(a_{\rm B}/a_0)$ is known as the Interpolating Function (IF), and we consider two families of IFs.  One is given by
\begin{equation}
  f_\nu(x) = \left(\frac{1}{2}+\sqrt{\frac{1}{4}+\frac{1}{x^{\nu}}}\right)^{1/\nu}, 
 \label{eq:IFnu}
\end{equation}
with $0<\nu\le 2$ which includes the simple ($\nu=1$) \citep{FB05} and the standard ($\nu=2$) \citep{Ken87} IFs. The other is given by
\begin{equation}
  f_\lambda(x) = \frac{1}{\left(1-{\rm e}^{-x^{\lambda/2}}\right)^{1/\lambda}} ,
 \label{eq:IFlam}
\end{equation}
with $0.3<\lambda<1.7$, which includes McGaugh's IF ($\lambda=1$) \citep{McG08}.

\section{Results} \label{sec:result}

The galaxies selected from the \atl\ sample are listed in Table~\ref{tab:chisq}. As described in Paper~I, photometric decomposition analyses detect no disks in any of these galaxies \citet{Kra13}, so we refer to them as pure-bulges. Two thirds (17/24) have low ellipticities $\varepsilon_{\rm e}\la 0.2$ within $R_{\rm e}$ (Table~\ref{tab:chisq}): the mean ellipticity for all 24 galaxies is $\langle\varepsilon_{\rm e}\rangle = 0.184$. In this sense these galaxies are referred to as nearly round (or spherical). Two thirds (16/24) are kinematic SRs within $R_{\rm e}$ dramatically different from the overall statistics of just 14\% (36/260) of SRs from the entire {\atl} sample. We analyze these 24 pure-bulges using the spherical Jeans equation, paying particular attention to the 16 SRs. 

\begin{deluxetable*}{cccccclcccccc}
\tablecaption{Observed properties of the {\atl} pure-bulge galaxies and quality of fit to the spherical model with the gNFW DM halo in the $\Lambda$CDM paradigm. Four different assumptions about the $M_\star/L$ gradient (parameterized by $K$) and velocity dispersion anisotropy ($\beta$) are considered: (a) $K=0$ (no $M_\star/L$ gradient) and $\beta={\rm constant}$; (b) $K=0$ and $\beta=\beta_{\rm gOM}(r)$; (c) $0<K<1.5$ and $\beta={\rm constant}$; (d) $0<K<1.5$ and $\beta=\beta_{\rm gOM}(r)$. \label{tab:chisq}}
\tabletypesize{\scriptsize}
\tablewidth{0pt}
\tablehead{
  \multicolumn{9}{r}{case:} & \colhead{(a)}  & \colhead{(b)}  &  \colhead{(c)}  &   \colhead{(d)} \\
  \cline{10-13}
\colhead{galaxy} & \colhead{$\varepsilon_{\rm e}$} & \colhead{$\log_{10}\sigma_{\rm e}$} & \colhead{$\log_{10}\sigma_{\rm e/8}$} & \colhead{$\eta_{<0.2 R_{\rm e}}$} & \colhead{$\eta_{\rm LW}$} & \colhead{kinematic feature} & \colhead{ $n$}  & \colhead{}  & \multicolumn{4}{c}{$\chi^2/N_{\rm dof}$ (``reduced chi-squared'')} \\
 \colhead{}  & \colhead{(1)}  & \colhead{(2)}  & \colhead{(3)}  & \colhead{(4)}  & \colhead{(5)} & \colhead{(6)}  & \colhead{(7)}   & \colhead{}  & \multicolumn{4}{c}{(8)}  
}
\startdata
  NGC 0661 & 0.306 & 2.251 & 2.279  & $-0.010 \pm  0.032$ &  $-0.031$   &  S: NRR/CRC & 20  &   &  1.2 & 1.2 & 1.3 &  1.4  \\
  NGC 1289  & 0.393 & 2.095 & 2.133 &  $+0.021 \pm  0.017$ &  $-0.042$    & S: NRR/CRC & 10  &   &  1.0 & 1.3 & 1.2 &  1.7  \\
  NGC 2695 & 0.293 & 2.257 & 2.342 &  $-0.247 \pm  0.023$ &  $-0.094$  &  F: RR & 23  &   & 23.6  & 2.9 & 15.2 &  1.3  \\
  NGC 3182 & 0.166 & 2.052 & 2.060 &  $-0.167 \pm  0.025$ &  $-0.009$ &  F: RR & 20  &   &  5.2  & 2.3  & 4.3 &  2.5  \\
  NGC 3193 & 0.129 & 2.252 & 2.299 &  $-0.083 \pm  0.008$ &  $-0.052$ &  F: RR & 24  &   & 2.0  & 2.1 & 2.0 &  2.2  \\
  NGC 3607 & 0.185 & 2.315 & 2.360 & $-0.020 \pm  0.004$ &  $-0.050$  &  F: RR & 34  &   & 1.3  & 0.8 & 1.7 &  0.9  \\
  NGC 4261 & 0.222 & 2.424 & 2.469 &  $-0.065 \pm  0.005$ &  $-0.050$ & S: NRR/NF & 32  &   & 3.8  & 1.2 & 1.4 &  1.2  \\
  NGC 4365 & 0.254 & 2.345 & 2.408 & $ -0.083 \pm  0.002$ &  $-0.070$ & S: NRR/KDC & 32  &   &10.1 & 5.5  & 5.7 &  4.4  \\
  NGC 4374 & 0.147 & 2.412 & 2.460 & $ -0.079 \pm  0.002$ &  $-0.053$ & S: NRR/LV & 33  &   & 2.8 & 2.6 & 2.6 &  2.7  \\
  NGC 4406 & 0.211 & 2.280 & 2.336 & $ -0.055 \pm  0.001$ &  $-0.062$ &  S: NRR/KDC & 34  &   & 7.2  & 0.9 & 1.4 &  0.9  \\
  NGC 4459 & 0.148 & 2.199 & 2.252 & $ -0.105 \pm  0.005$ &  $-0.059$  &  F: RR & 30  &   & 7.2  & 6.3 & 2.4 & 2.4  \\
  NGC 4472 & 0.172 & 2.398 & 2.460 & $ -0.050 \pm  0.001$ &  $-0.069$  &  S: NRR/CRC & 31  &   & 1.7  & 1.4  & 1.4 &  1.5  \\
  NGC 4486 & 0.037 & 2.422 & 2.497 & $ -0.138 \pm  0.002$ &  $-0.083$ &  S: NRR/LV & 34  &   & 13.9 & 6.0 & 2.9 &  2.6  \\
  NGC 4636 & 0.094 & 2.259 & 2.300 & $ -0.023 \pm  0.002$ &  $-0.045$  &  S: NRR/LV & 33  &   & 1.7  & 1.5 & 1.7 &  1.5  \\
  NGC 4753 & 0.213 & 2.241 & 2.263 & $ -0.023 \pm  0.002$ &  $-0.045$  &  S: NRR/LV & 34  &   & 9.7  & 3.7  & 9.3 &  2.7  \\
  NGC 5322 & 0.307 & 2.351 & 2.395 & $ +0.012 \pm  0.006$ &  $-0.049$ &  S: NRR/CRC & 26  &   & 1.9  & 1.2 & 1.9 &  1.3  \\
  NGC 5481 & 0.214 & 2.085 & 2.174 & $ -0.263 \pm  0.032$ &  $-0.099$ &  S: NRR/KDC & 11  &   & 2.7  & 2.8  & 1.9 &  2.6  \\
  NGC 5485  & 0.171 & 2.223 & 2.253 & $ -0.091 \pm  0.013$ &  $-0.033$ &  F: NRR/NF & 19  &   & 4.6  & 2.6 & 3.2 &  2.5  \\
  NGC 5557 & 0.169 & 2.306 & 2.406 & $ -0.185 \pm  0.015$ &  $-0.111$  &  S: NRR/NF & 32  &   & 2.7  & 1.5  & 1.3 &  1.4  \\
  NGC 5631  & 0.127 & 2.176 & 2.207 & $ -0.097 \pm  0.013$ &  $-0.034$ &  S: NRR/KDC & 24  &   & 2.7  & 1.5  & 2.2  &  1.7  \\
  NGC 5831  & 0.136 & 2.158 & 2.220 & $ -0.058 \pm  0.007$ &  $-0.069$ &  S: NRR/KDC & 23  &   & 2.4  & 2.1 & 1.9 &  2.1  \\
  NGC 5846  & 0.062 & 2.349 & 2.365 & $ -0.033 \pm  0.003$ &  $-0.018$ &  S: NRR/LV & 31  &   & 2.6   & 2.6 & 2.3 &  2.3  \\
  NGC 5869  & 0.245 & 2.224 & 2.260 & $ -0.146 \pm  0.011$ &  $-0.040$ &  F: RR & 19  &   & 1.9   & 2.1  & 2.0  &  2.2  \\
  NGC 6703  & 0.019 & 2.178 & 2.260 & $ -0.072 \pm  0.009$ &  $-0.091$ &  S: NRR/LV & 30 &  & 2.8   & 1.4  & 2.4  &  1.4  \\
  \hline
\multicolumn{9}{r}{$N_{\rm dof}=$}  & \colhead{ $n-3$}  &  \colhead{$n-5$}  &  \colhead{$n-4$}   &  \colhead{$n-6$} \\
\enddata
\tablecomments{(1) Ellipticity of the observed surface brightness distribution within $R_{\rm e}$ taken from \citet{Cap13}. (2) $\sigma_{\rm e}$ = the effective velocity dispersion, i.e.\ the light-weighted {\siglos} within $R_{\rm e}$, in \kms taken from \citet{Cap13}. (3) $\sigma_{\rm e/8}$ = the central velocity dispersion, i.e.\ the light-weighted {\siglos} within $R_{\rm e/8}$, in \kms taken from \citet{Cap13b}. (4) $\eta_{<0.2 R_{\rm e}}$ = the logarithmic slope of the {\siglos} profile within $R<0.2 R_{\rm e}$. (5) $\eta_{\rm LW}$ = the logarithmic slope between $R_{\rm e}/8$ and $R_{\rm e}$ of the light-weighted (and integrated) values of {\siglos} as given in this table. (6) Slow(S)/fast(F) rotator identifications within $R_{\rm e}$ come from \citet{Ems11}. Kinematic features come from \citet{Kra11}. The acronyms mean the following (see the text for further details): RR - regular rotator, NRR - non-regular rotator, KDC - kinematically distinct core, CRC - counter rotating core (which is a special case of KDC),  LV - low-level (rotation) velocity, NF - no feature. Only one NRR (NGC 5485) is classified as a fast rotator. (7) $n$ = the number of the measured {\siglos} values, i.e.\ the number of the bins in $R$. (8) Minimum values of the ``reduced $\chi^2$'', i.e.\ $\chi^2$ per the number of the degree of freedom ($N_{\rm dof}$) for the four different cases considered. $N_{\rm dof}$ for each case is indicated in the las row.  A value of $\chi^2/N_{\rm dof} = 1$ means that the measured $\sigma_{\rm los}(R)$ values are matched at $\sim 1\sigma$,  4 means $\sim 2\sigma$, etc. For the most realistic case of (d) all galaxies but NGC 4365 have $\chi^2/N_{\rm dof} \la 2.7$ meaning that the measured $\sigma_{\rm los}(R)$ are matched at $\sim 1.6\sigma$.}
\end{deluxetable*}

In what follows, we study four different assumptions about the $M_\star/L$ gradient (parameterized by $K$) and velocity dispersion anisotropy ($\beta$):

(a) $K=0$ (no $M_\star/L$ gradient) and $\beta={\rm constant}$; 

(b) $K=0$ and $\beta=\beta_{\rm gOM}(r)$; 

(c) $0<K<1.5$ and $\beta={\rm constant}$; 

(d) $0<K<1.5$ and $\beta=\beta_{\rm gOM}(r)$.\\
Thus, for our most general case (d), a spherical model with a gNFW DM halo (Equation~(\ref{eq:gNFW})) has six free parameters, i.e.\ $\Upsilon_{\star 0}$, $K$, $\alpha$, $\beta_0$, $\beta_{\infty}$, and $r_a$ as well as the following constrained parameters: $M_{200}$ (halo mass), $c=r_{200}/r_s$ (halo concentration), and $M_{\rm BH}$ (black hole mass) (see Paper~I and \citealt{CBS18b} for further details).  The number of free parameters is reduced for cases (a) and (b) where we set $K=0$, and for (a) and (c) when we set $\beta = $ constant.

In each case, the model is fitted to the line-of-sight velocity dispersion\footnote{In Paper~I we referred to this using the acronym ``LOSVD'' for ``line-of-sight velocity dispersion''. } radial profile, \siglos, constructed from the velocity dispersion map. We do this by minimizing 
\begin{equation}
  \chi^2 \equiv \sum_{i=1}^{n}\left(\frac{\sigma_{\rm los}^{\rm obs}(R_i)-\sigma_{\rm los}^{\rm mod}(R_i)}{s_i}\right)^2,
  \label{eq:chisq}
\end{equation}
where $\sigma_{\rm los}^{\rm obs}(R_i)$ and $\sigma_{\rm los}^{\rm mod}(R_i)$ refer to the observed and predicted velocity dispersions at the projected radii $R_i$, $s_i$ are the observational uncertainties, and $n$ is the number of radial bins as given in Table~\ref{tab:chisq}. The number of the degree of freedom ($N_{\rm dof}$) for each model is then given by $N_{\rm dof} = n- N_{\rm free}$ where $N_{\rm free}$ is the number of free parameters ranging from 3 - 6.  We call $\bar{\chi}^2\equiv\chi^2/N_{\rm dof}$ the ``reduced $\chi^2$''.

For each galaxy we produce a set of 400 MC models for the case of $K=0$ (constant $M_\star/L$) and a set of 800 MC models when we allow $0< K <1.5$.  MC models for each galaxy are produced iteratively based on Equation~(\ref{eq:chisq}) from the prior ranges of the model parameters. The details of this procedure can be found in Paper~I and \citet{CBS18b}. \citet{CBS18b} further describes the distribution of $\bar{\chi}^2$ in the MC set and the fitted {\siglos} profiles. Each MC set provides posterior probability density functions (PDFs) of the free parameters. Note that, although we are using the gOM model (Equation~(\ref{eq:gOM})) to describe the anisotropy profile, we also produce results for the case of constant anisotropies to quantify the effects of varying anisotropies and provide a direct comparison with relevant previous literature.

\subsection{Quality of fits}
Table~\ref{tab:chisq} summarizes the fit qualities of the aforementioned four different cases under the $\Lambda$CDM paradigm with the gNFW DM halo model (Equation~(\ref{eq:gNFW})).  For the most general and realistic case (d) -- the gOM model with $K$ allowed to vary within $0< K <1.5$ -- the measured line-of-sight velocity dispersions of all 24 galaxies (with the exception of NGC~4365) are matched within $\sim 1.6\sigma$ by the best-fitting models. In other words, the reduced $\chi^2$ has minimum values $\bar{\chi}^2\le 2.7$ for the 23 galaxies with an average of  $\langle\bar{\chi}^2\rangle=1.87$ while NGC~4365 has $\bar{\chi}^2=4.4$ (these values are somewhat different from, and are meant to replace, the values given in Figure~6 of Paper~I because we have revised MC samples and corrected $N_{\rm dof}$ here).

\begin{figure*} 

  \vspace{1em}
\hspace{1em}   \includegraphics[angle=-90,origin=c,scale=0.4]{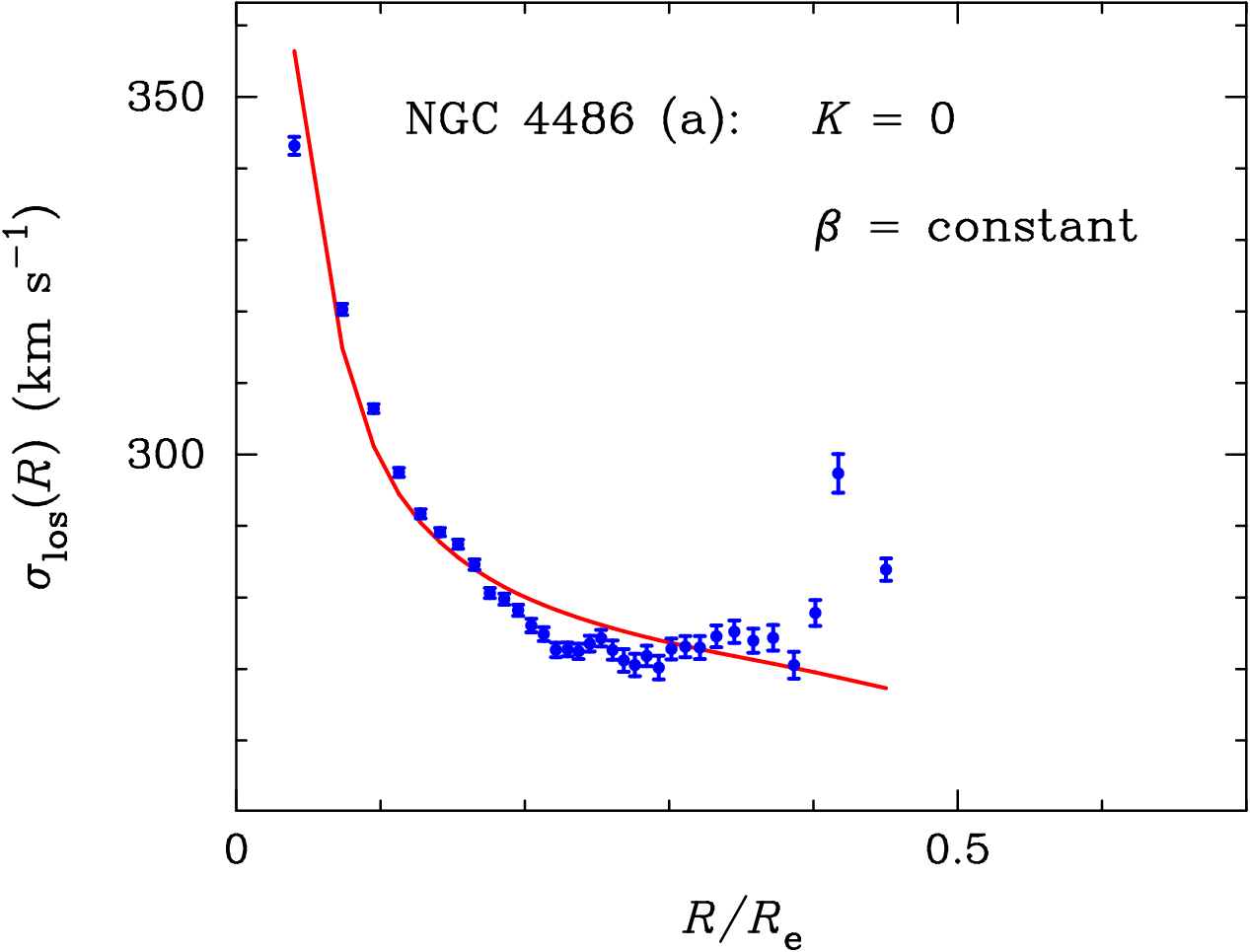}
\hspace{0em}  \includegraphics[angle=-90,origin=c,scale=0.4]{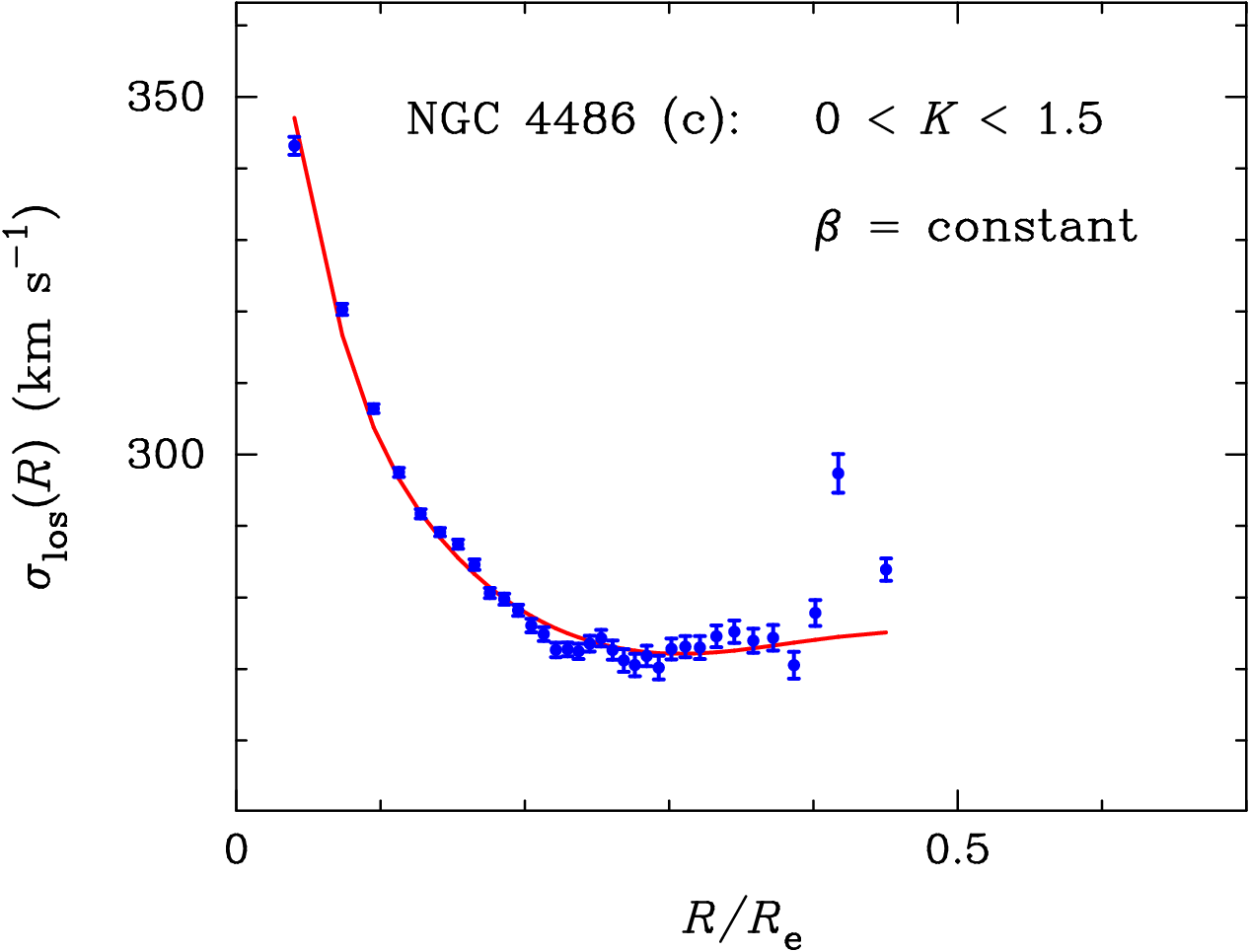}

\hspace{1em}   \includegraphics[angle=-90,origin=c,scale=0.4]{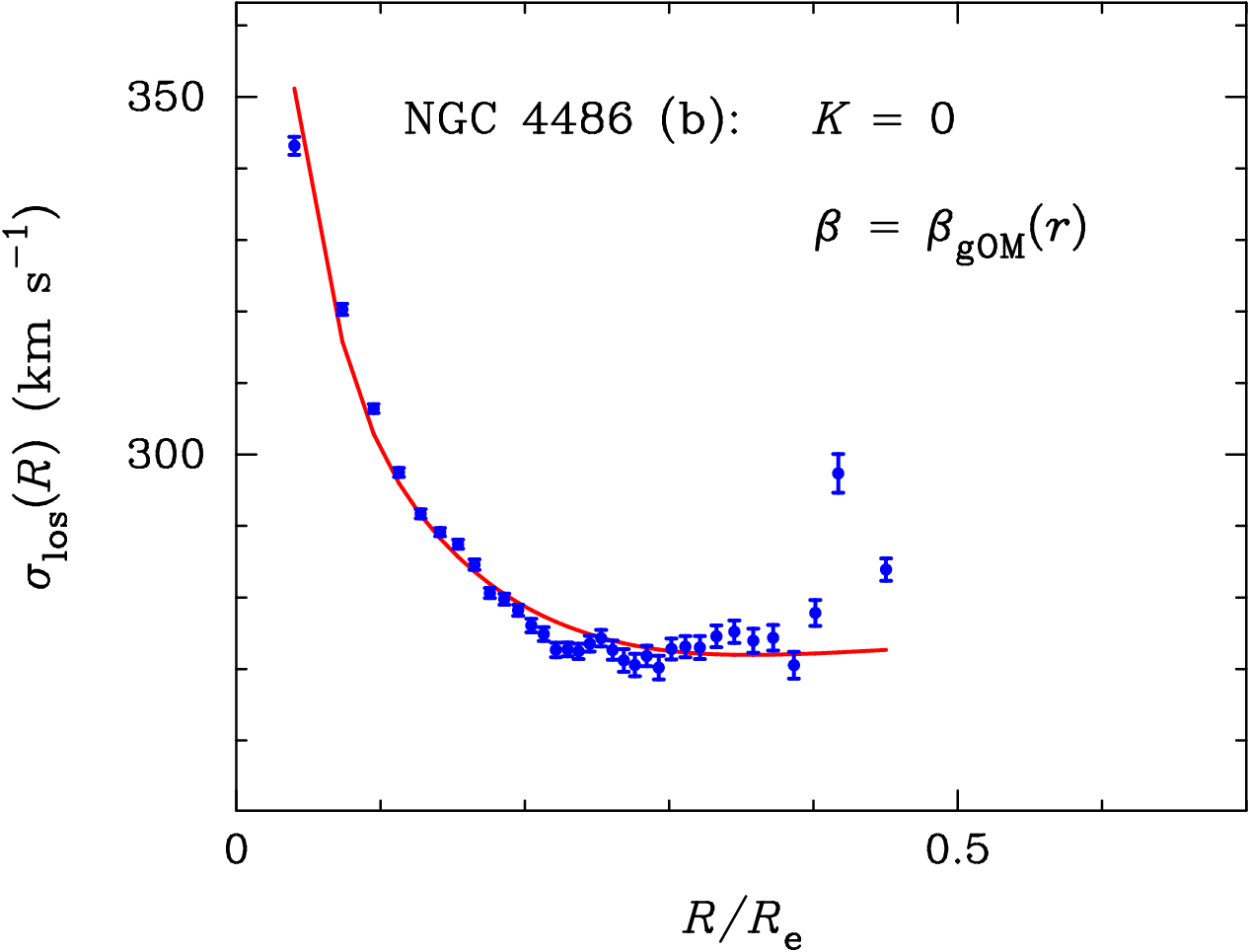}
\hspace{0em}  \includegraphics[angle=-90,origin=c,scale=0.4]{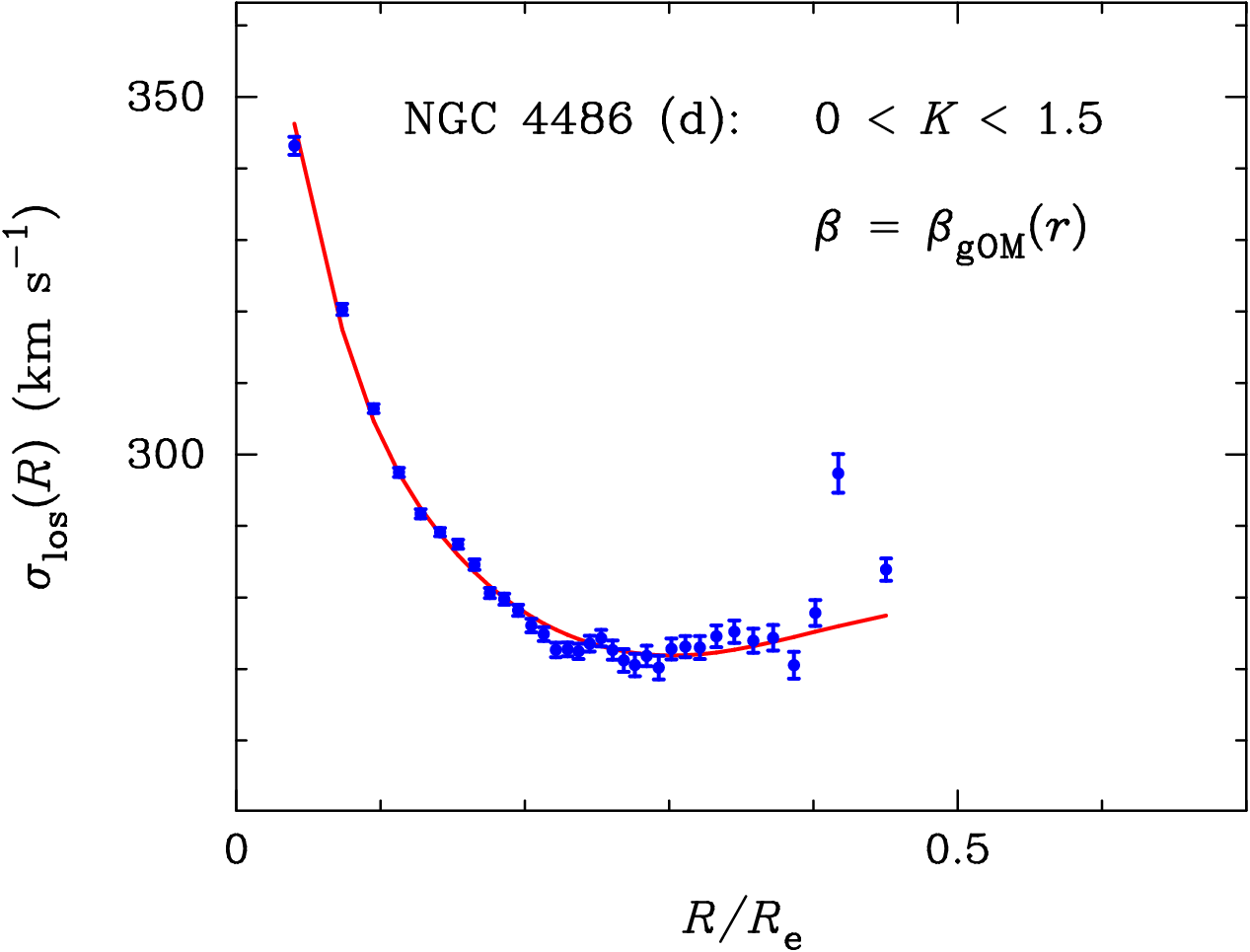}

  \vspace{-2em}
\caption{Comparison of the observed {\siglos} profile of NGC 4486 with the best-fit model predictions for the four different cases of Table~\ref{tab:chisq}. Note that successful fits can be achieved only for the cases of (c) and (d) for which an $M_\star/L$ radial gradient is allowed.   \label{fig:VP4486}}
\end{figure*}

However, for the simplest, most restrictive case (a) -- spatially constant anisotropy and no $M_\star/L$ gradient -- which has been sometimes adopted in the literature, 8 galaxies have unacceptably large $\bar{\chi}^2> 4$, and only 15 have $\bar{\chi}^2 \la 2.8$. If an $M_\star/L$ gradient is allowed with $0< K <1.5$ for the constant anisotropy model, then 4 galaxies (NGC 2695, 3182, 4365, 4753) have $\bar{\chi}^2> 4$. If the varied anisotropy model is used with $K=0$, then 3 galaxies (NGC 4365, 4459, 4486) have $\bar{\chi}^2> 4$. For these three galaxies the fit is clearly improved if we allow $K>0$. Figure~\ref{fig:VP4486} illustrates this point for NGC~4486 (M87). This means that, for these galaxies, reasonable gradients in $M_\star/L$ suffice to explain the current data without invoking drastically varying anisotropies (see, e.g., the classical discussion in \citet{BM82}). In particular, our conclusion for NGC 4486 agrees with \citet{Old18}. For the others, the fit for $0<K<1.5$ is as good as or better than for $K=0$. What is more significant is that the PDFs of $K$ do not in general prefer $K=0$, meaning that $K=0$ should not be presumed unless independent observational constraints require it. Therefore, in inferring the orbital anisotropies $M_\star/L$ gradients should be allowed.

\begin{figure*} 
  \vspace{-3em}
  \hspace{8em}  \includegraphics[angle=-90,origin=c,scale=0.55]{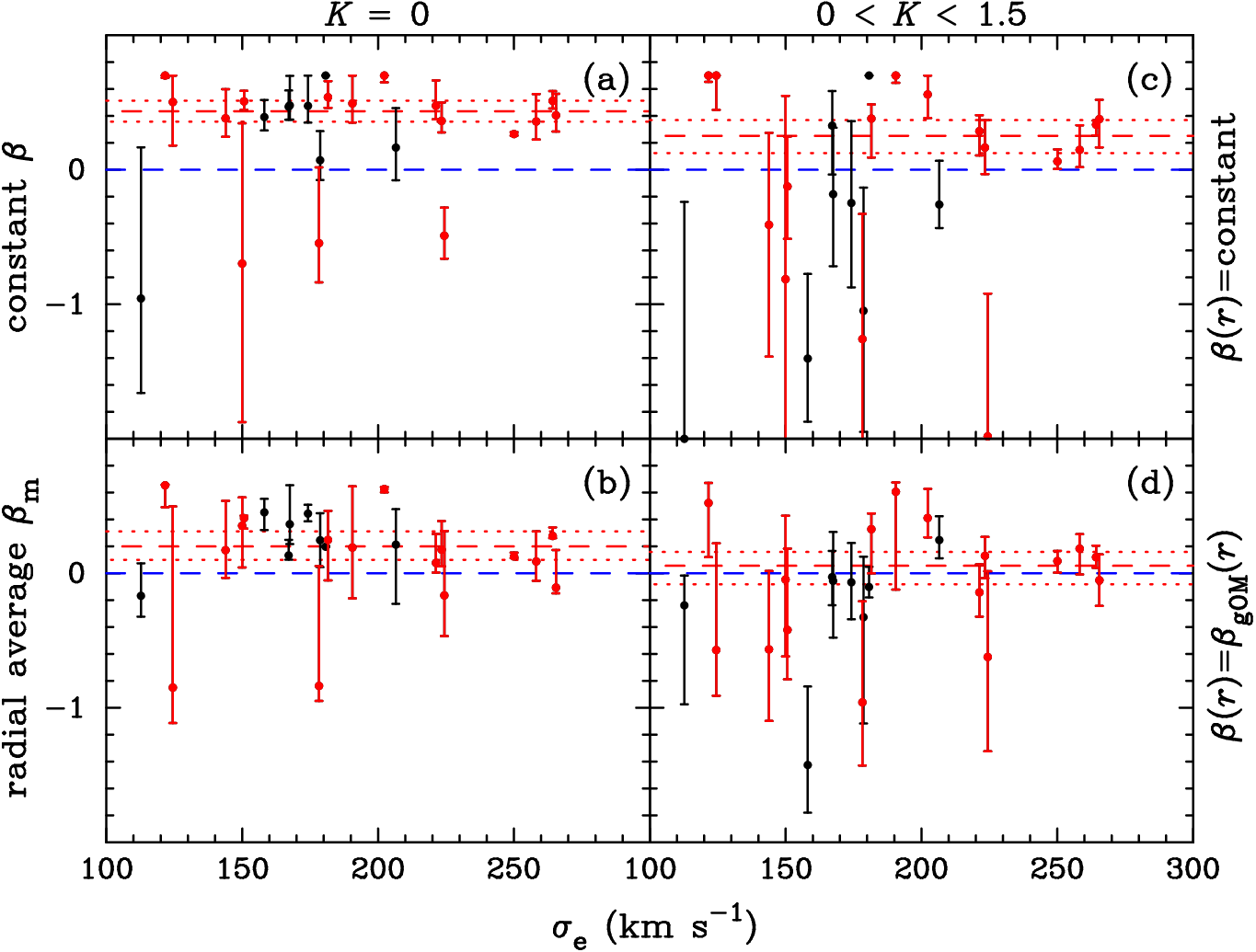}

  \vspace{-3em}
\hspace{8em}  \includegraphics[angle=-90,origin=c,scale=0.55]{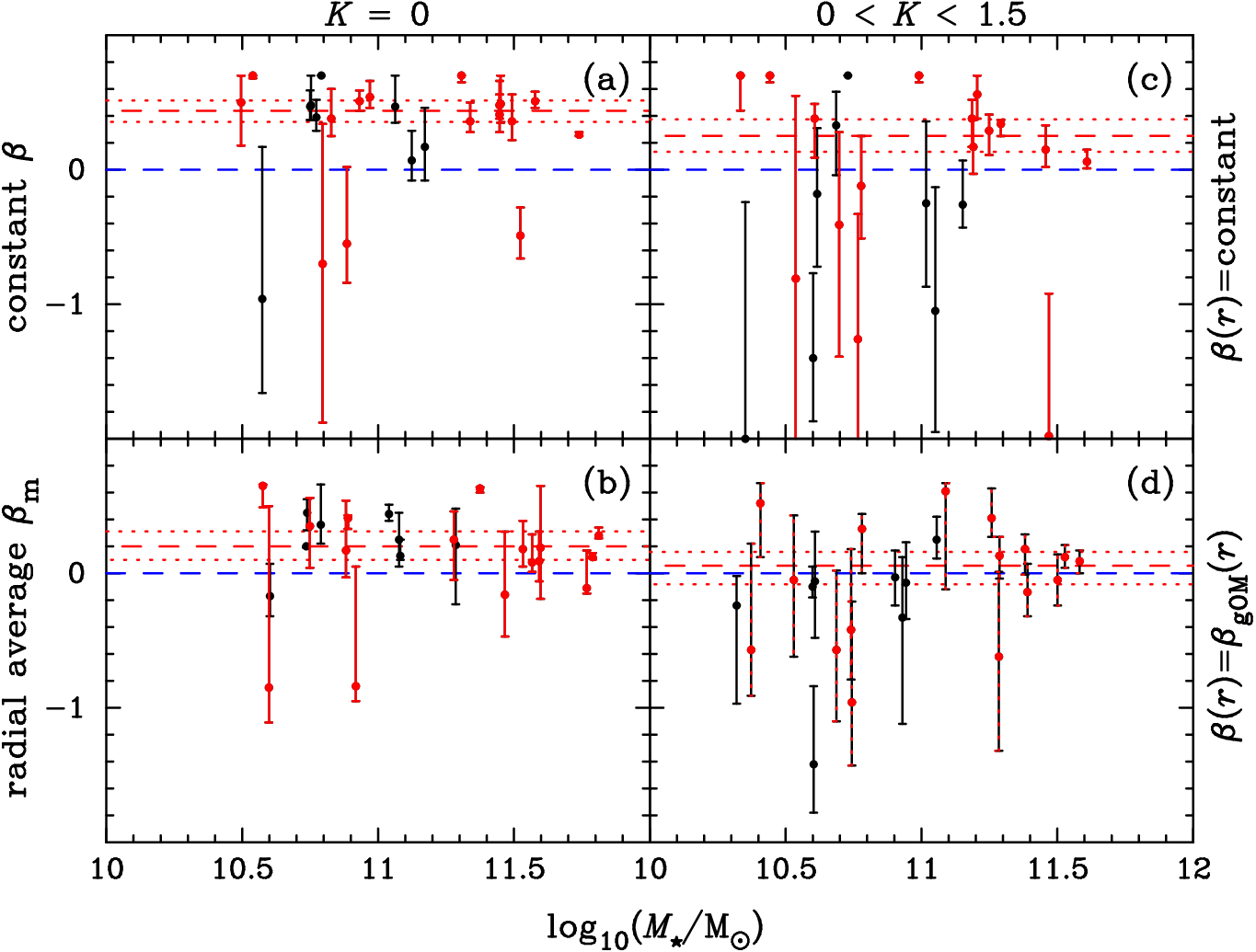}

  \vspace{-4em}
\caption{Fitted anisotropies with respect to the effective velocity dispersion $\sigma_{\rm e}$ (upper) and the fitted stellar mass $M_\star$ (lower) of the 24 {\atl} pure-bulges. Red points represent 16 SSBGs. Red dashed line is the median estimated from the composite PDFs whose examples are shown in Figure~\ref{fig:pdfbetm}. Red dotted lines are the 68\% uncertainties of the medians. Four different cases of Table~\ref{tab:chisq} are considered. The numerical values are provided in Table~\ref{tab:beta} and Table~\ref{tab:MsL}. \label{fig:bet4gNFW}}
\end{figure*}

\subsection{Inferred anisotropy}
\begin{figure*} 
  \vspace{1em}
  
  \hspace{-5em}
\includegraphics[scale=0.65]{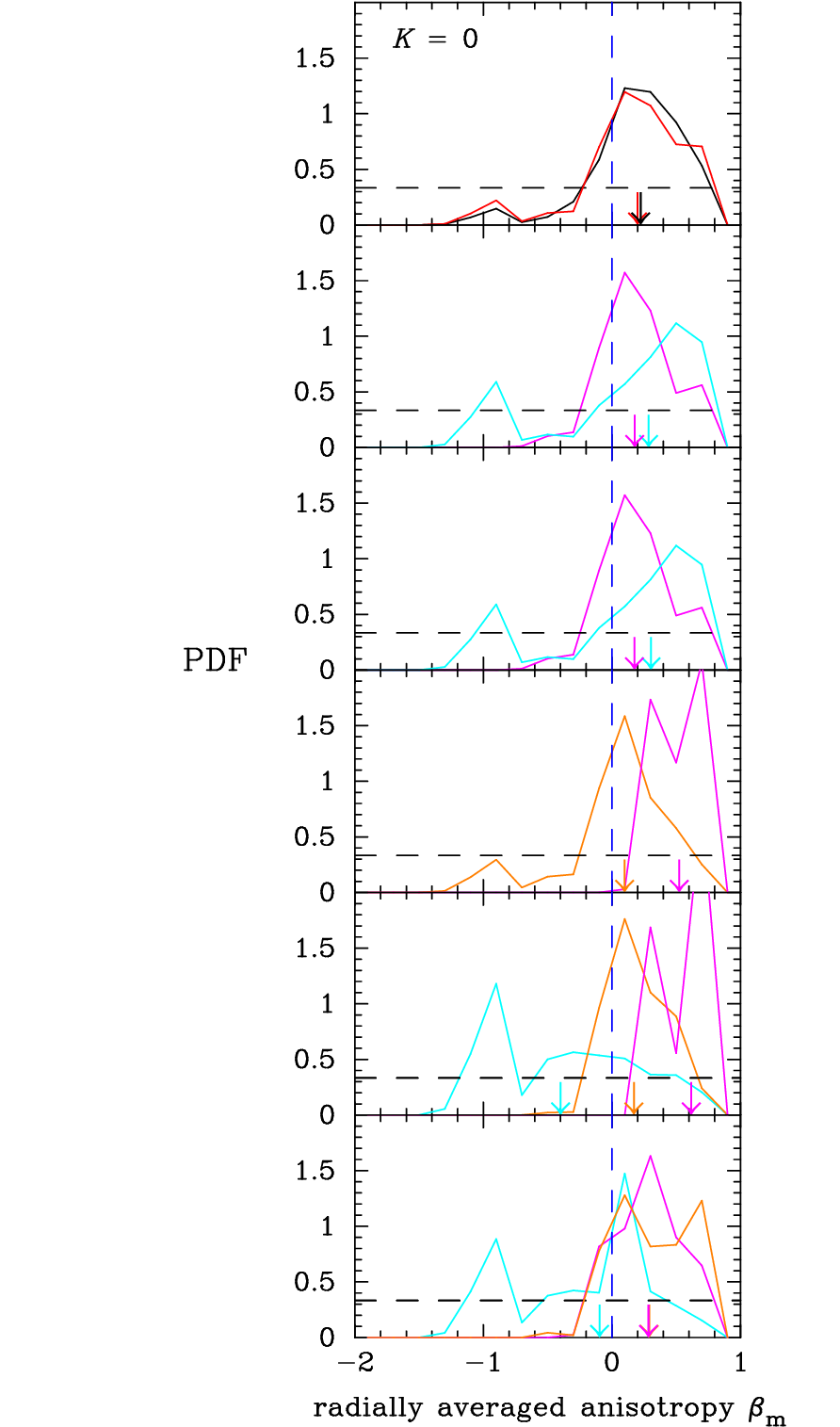}\includegraphics[scale=0.65]{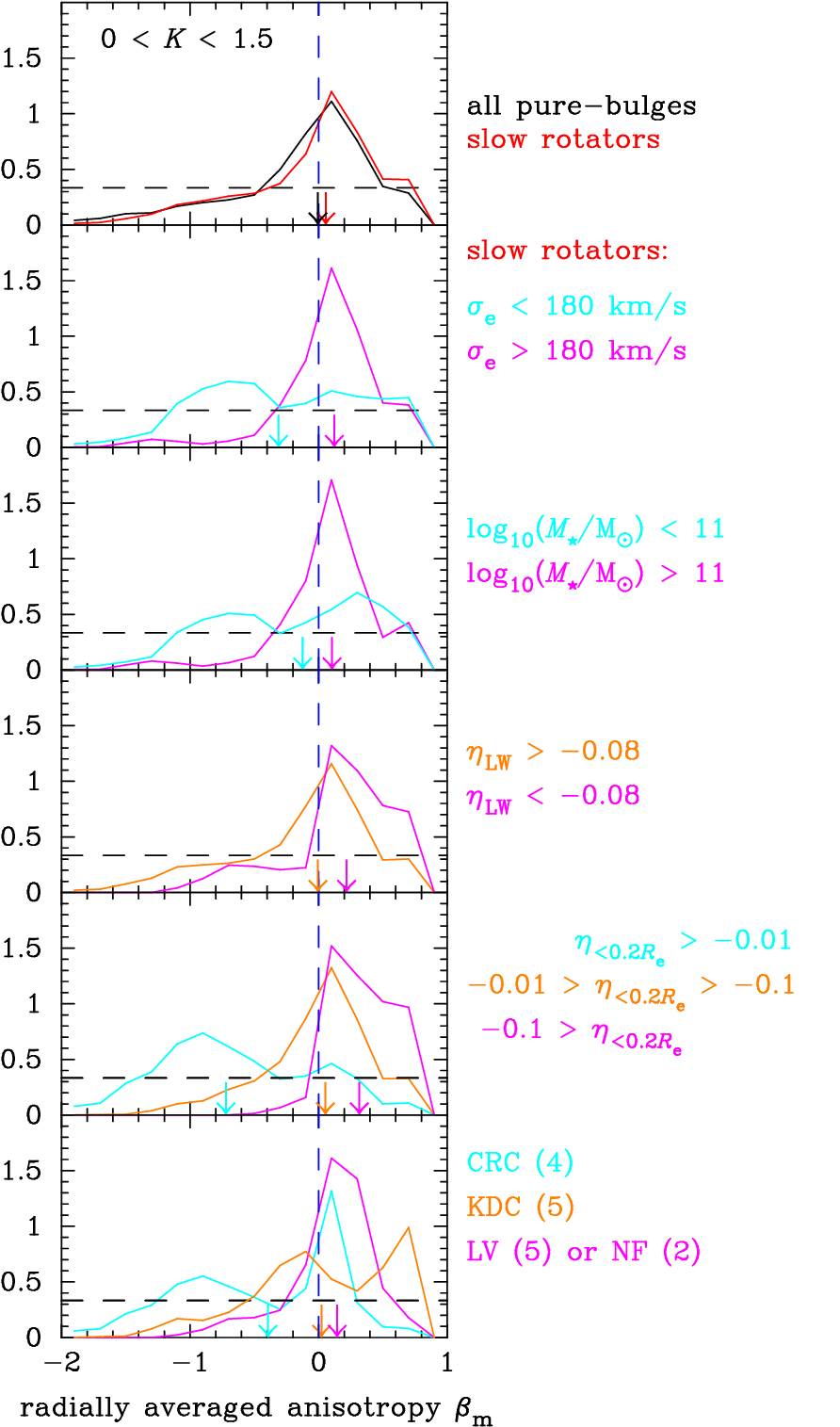}
\caption{PDFs of $\beta_{\rm m}$ for the cases of $K=0$ and $0<K<1.5$ with the gOM anisotropy model: cases (b) and (d) of Table~\ref{tab:chisq}. The top panels show the PDFs for all 24 pure-bulge galaxies (black) and 16 slow rotators (red). The other panels show the PDFs for sub-samples of slow rotators split respectively by $\sigma_{\rm e}$ (effective velocity dispersion), $M_\star$ (fitted stellar mass), $\eta_{\rm LW}$ (logarithmic slope of the light-weighted line-of-sight velocity dispersion $\langle\sigma_{\rm los}\rangle(R)$ between $R_{\rm e}/8$ and $R_{\rm e}$), $\eta_{<0.2 R_{\rm e}}$ (logarithmic slope of the {\siglos} profile for $R <0.2 R_{\rm e}$) and kinematic features (see Table~\ref{tab:chisq}). The downward pointing arrows indicate the medians in the PDFs. \label{fig:pdfbetm}}
\end{figure*}

Figure~\ref{fig:bet4gNFW} exhibits the constrained anisotropy values with respect to $\sigma_{\rm e}$, the effective velocity dispersion within $R_{\rm e}$ from {\atl} \citep{Cap13}, and the constrained stellar mass, $M_\star$ (corresponding to the MGE light distribution: see Paper~I) for the four different cases of Table~\ref{tab:chisq}. Table~\ref{tab:beta} gives the numerical values of the fitted anisotropies while Table~\ref{tab:MsL} gives $M_\star/L$ with respect to the SDSS $r$-band luminosity \citep{Cap13}. The median value for each case is derived from the composite probability density function (PDF) of the individual PDFs with a uniform weighting. Its statistical uncertainty is estimated from a Monte Carlo method using the composite PDF. For cases (b) and (d), the composite PDFs are displayed in Figure~\ref{fig:pdfbetm}. For these cases with the gOM model we consider a radially averaged value given by
\begin{equation}
  \beta_{\rm m} \equiv \int_0^{r_{\rm max}} \beta(r) dr / r_{\rm max},
 \label{eq:betm}
\end{equation}
where $r_{\rm max}$ is the maximum radius of the constructed {\siglos} profile ($\la R_{\rm e}$).

We see that the inferred anisotropies depend on the assumption on the anisotropy profile and $M_\star/L$ gradient. For the cases of constant $M_\star/L$ ($K=0$) the median anisotropies are radially biased with $\langle\beta\rangle = 0.43\pm 0.08$ (for the constant anisotropy: case (a)) or $\langle\beta_{\rm m}\rangle = 0.20^{+0.11}_{-0.10}$ (for the gOM anisotropy: case (b)). The former result (case (a)) agrees well with an estimate $0.45\pm 0.25$ from the combined analysis of lensing and stellar dynamics by \citet{Koo09} under the same assumption. The latter result (case (b)) also agrees well with the literature results. \citet{Ger01} obtain a median value of $\beta_{\rm m}\approx 0.2$ within $R_{\rm e}$ (their Figure~5) through a orbit superposition modeling of 21 nearly round and slowly rotating galaxies assuming spherical galaxy models. \citet{Cap07} present various anisotropy parameters based on axisymmetric dynamical modeling of 24 SAURON ETGs. For three SRs in common with our galaxy sample, the spherical anisotropy in Table~2 of \citet{Cap07} is $\beta_{\rm m}=0.11$ (NGC~4374), $\beta_{\rm m}=0.24$ (NGC~4486), and $\beta_{\rm m}=0.17$ (NGC~5846).  All are in good agreement with our results for the case (b) shown in Table~\ref{tab:beta}.

Figures~\ref{fig:bet4gNFW} and \ref{fig:pdfbetm} indicate that for $0< K <1.5$ the anisotropy distribution has a larger spread compared with the case for $K=0$. This appears to be a consequence of the shift of the anisotropy values for selective galaxies to lower values for $0< K <1.5$ from the case for $K=0$. The second panels from the top in Figure~\ref{fig:pdfbetm} show that the PDF for the lower-$\sigma_{\rm e}$ galaxies for $0< K <1.5$ has a larger spread than that for $K=0$. On the other hand, the PDF for higher-$\sigma_{\rm e}$ galaxies does not show a significant shift between the two cases.

\begin{figure*}[ht!] 
\begin{center}
\setlength{\unitlength}{1cm}
\begin{picture}(17,17)(0,0)
\put(1.,-1.2){\includegraphics{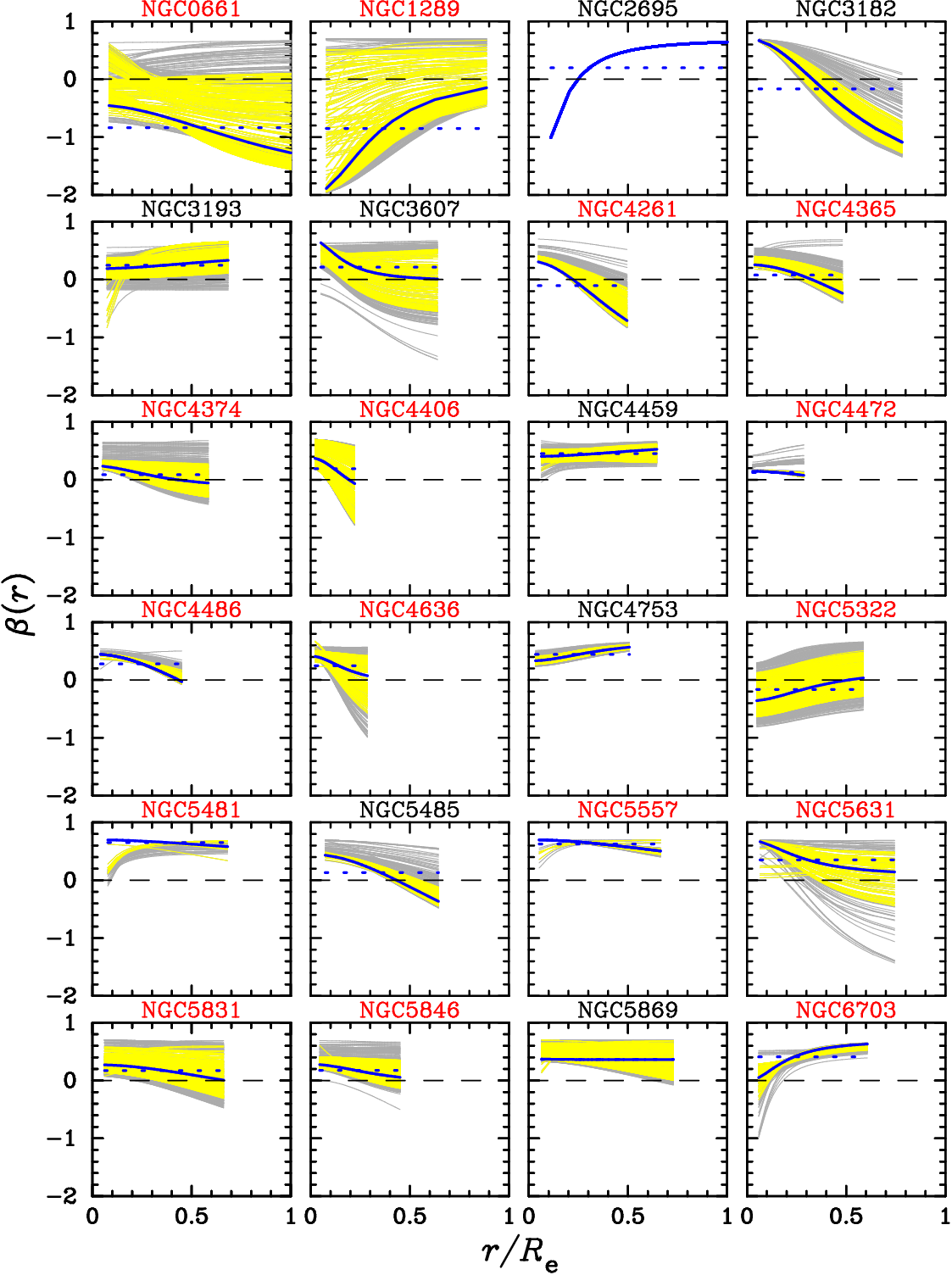}}
\end{picture}
\caption{Constrained anisotropy profiles for case (b) of Table~\ref{tab:chisq} (i.e.\ for $K=0$). All MC models satisfying $\bar{\chi}^2 < 2 \bar{\chi}^2_{\rm min}$ are shown (cf. Paper I). Yellow curves represent only 68\% of the models from the median denoted by the blue curve. The blue dotted line shows $\beta_{\rm m}$ obtained from this blue curve from Equation~(\ref{eq:betm}). For slow rotators, the galaxy ID is shown in red. \label{fig:betgOMcML}}
\end{center}
\end{figure*}

Comparison of case~(b) with case~(a) shows that for many galaxies with unacceptable fits when the anisotropy is assumed constant (see Table~\ref{tab:chisq}), the introduction of radially varying anisotropies can improve the fit dramatically, qualitatively consistent with dynamical modeling results (e.g.\ \citealt{Ger01,Geb03,Tho07}). The anisotropy profile $\beta(r)$ inferred for $K=0$ is shown in Figure~\ref{fig:betgOMcML}. While our smooth model may not capture fine details that might be recovered from orbit-superposition modeling (e.g.\ \citealt{Ger01,Geb03,Tho07}), it appears to capture overall radial trends within the relatively central optical regions. We find both radially increasing and declining $\beta(r)$. Note, however, for several galaxies (NGC 4365, 4459, 4486, and 4753), if we fix $K=0$, then the fit quality is still not good even with the radially varying anisotropy model of $\beta_{\rm gOM}(r)$. 

\begin{figure*}[ht!] 
\begin{center}
\setlength{\unitlength}{1cm}
\begin{picture}(17,17)(0,0)
\put(1.,-1.2){\includegraphics{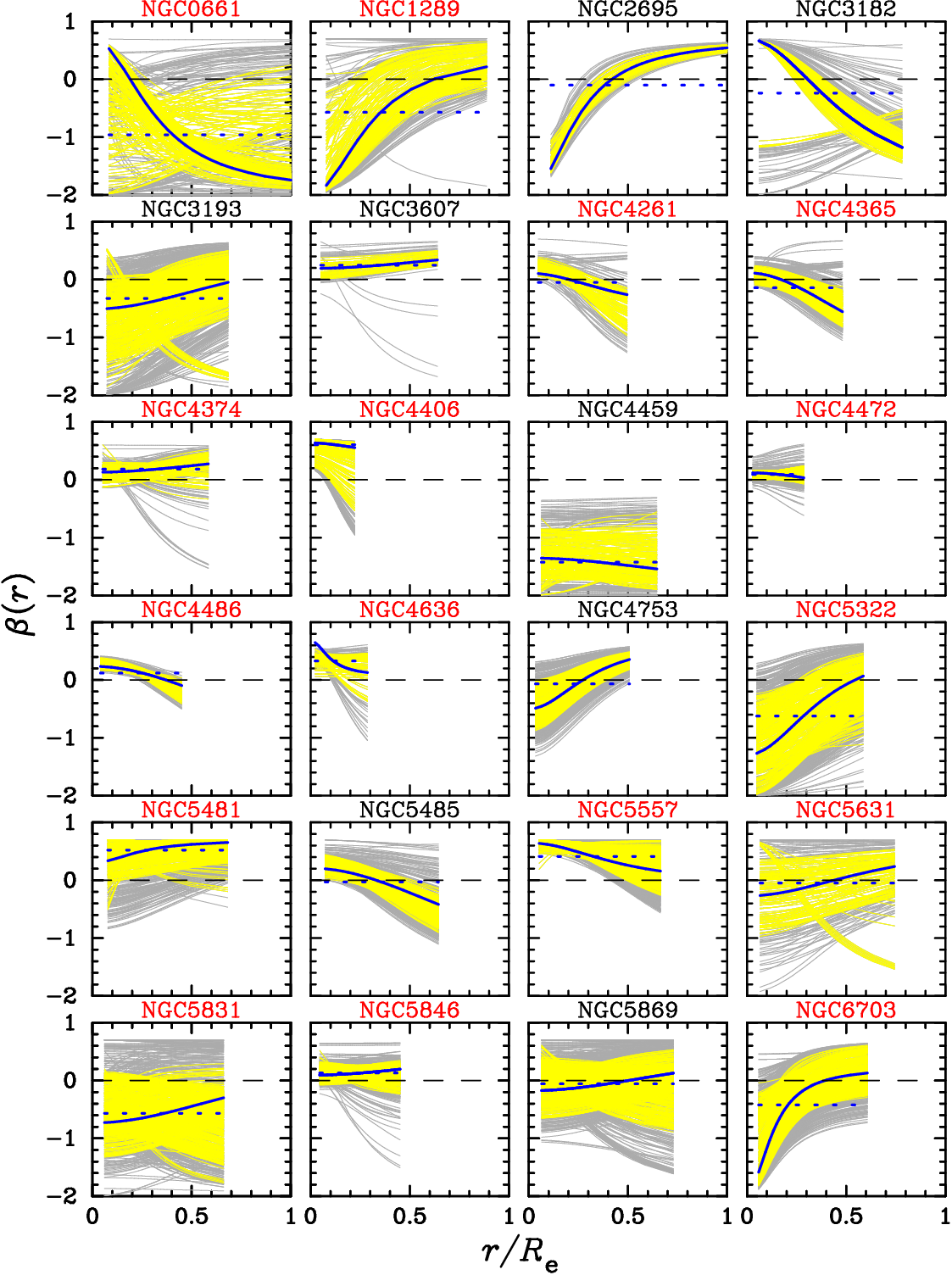}}
\end{picture}
\caption{Same as Figure~\ref{fig:betgOMcML}, but with the constraint $0<K<1.5$; i.e., case (d) of Table~\ref{tab:chisq} . \label{fig:betgOM}}
\end{center}
\end{figure*}

When an $M_\star/L$ gradient is allowed, the fits are improved, sometimes dramatically. Interestingly, even for the constant anisotropy model, the improvement is dramatic except for 4 galaxies (NGC 2695, 3182, 4365, 4753). This implies significant degeneracies between $K$ and anisotropy shape. Nevertheless, to obtain successful fits for {\it all} galaxies we require both $M_\star/L$ radial gradient and radial variations in $\beta$. With the constraint $0 < K < 1.5$ we have $\langle\beta_{\rm m}\rangle = 0.25^{+0.12}_{-0.13}$ (for the constant anisotropy: case (c)) or $\langle\beta_{\rm m}\rangle = 0.06^{+0.10}_{-0.12}$ (for the gOM anisotropy: case (d)). Compared with the corresponding cases with $K=0$, the median anisotropies are reduced by $\Delta \langle \beta_{\rm m} \rangle \approx -0.2$. Figure~\ref{fig:betgOM} shows the anisotropy profiles $\beta(r)$ for $0< K <1.5$.

Replacing the gNFW profile the Einasto form (Equation~(\ref{eq:Ein})) yields similarly good fits (i.e. $\bar{\chi}^2$ values) and anisotropy profiles, so we do not exhibit them. When the MOND models (Equations~(\ref{eq:IFnu}) and (\ref{eq:IFlam})) are used, we also obtain similar $\bar{\chi}^2$ and $\beta$. This means that our results are robust with respect to model choices in both $\Lambda$CDM and MOND paradigms. 

\subsection{Anti-correlation between $\beta$ and $M_\star/L$ gradient}
\begin{figure*} 
\begin{center}
\setlength{\unitlength}{1cm}
\begin{picture}(13,12)(0,0)
\put(-2.,13.4){\includegraphics{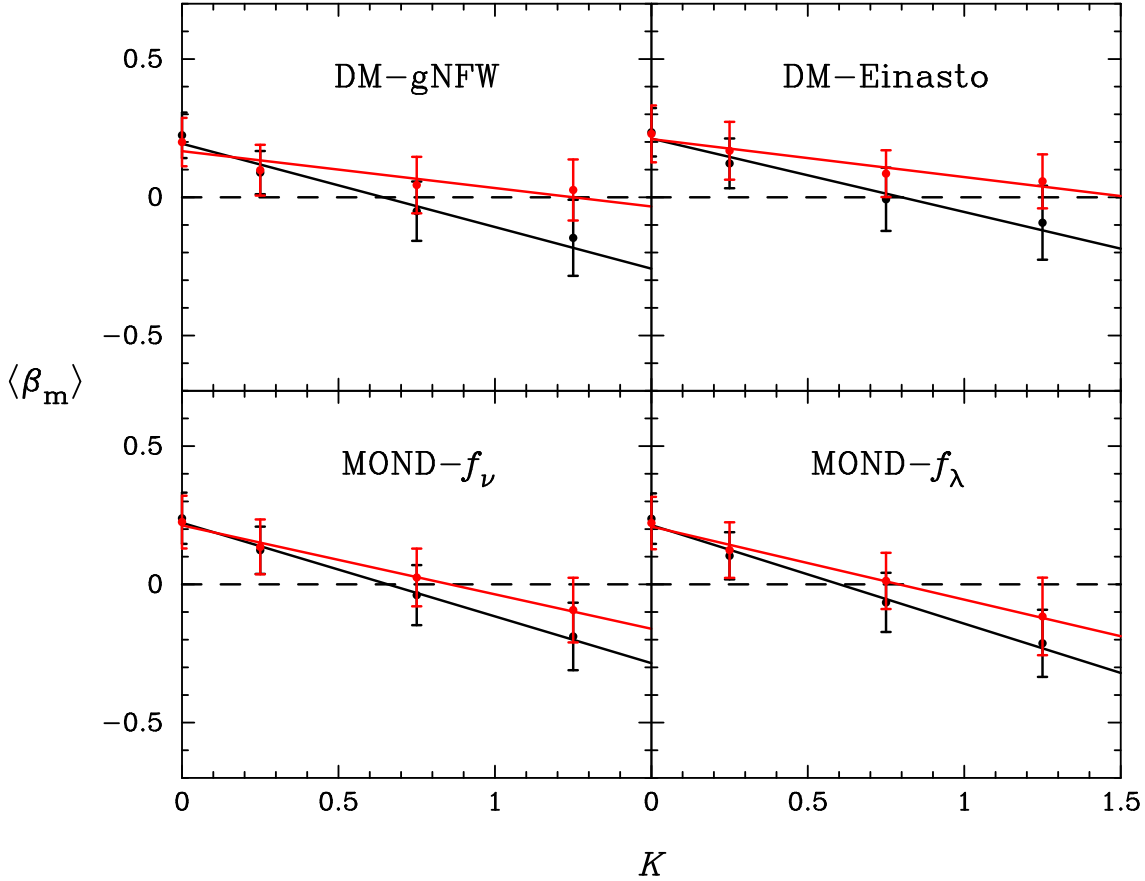}}
\end{picture}
\caption{Median fitted anisotropies with respect to $K$ with the gOM anisotropy model given by Equation~(\ref{eq:gOM}). Two models (Equations~\ref{eq:gNFW} and \ref{eq:Ein}) of the $\Lambda$CDM (upper panels) and two models (Equations~\ref{eq:IFnu} and \ref{eq:IFlam}) of the MOND (lower panels) are considered. Black points and lines represent all 24 pure-bulges while red ones represent the 16 slow rotators. \label{fig:betmK}}
\end{center}
\end{figure*}

Strikingly, for the most general case (d), isotropic velocity dispersions ($\beta_{\rm m}=0$) are preferred. To understand why, we split the MC models for the case of radially varying anisotropy into four bins in $K$ including the special case of $K=0$. We then analyze the MC models in each bin and obtain the anisotropies. The results are displayed in Figure~\ref{fig:betmK}, which shows a clear trend for $\langle\beta_{\rm m}\rangle$ to decrease as $K$ increases.  Combining the two $\Lambda$CDM results, i.e.\ for the gNFW and Einasto profiles, we find
\begin{equation}
  \langle\beta_{\rm m}\rangle \approx a + b K,
\end{equation}
with $a=0.19\pm 0.05$ and $b=-0.13\pm 0.07$. We obtain similar coefficients $a=0.21\pm 0.05$ and $b=-0.26\pm 0.08$ from the MOND results. Clearly, $\beta>0$ is obtained only if $K \approx 0$, although the bias towards more radial orbits may not be large.  As $K$ gets larger than $\sim 0.5$, $\beta=0$ starts to be preferred while at large $K \ga 1$ a tangential bias ($\beta<0$) is preferred. As shown in Paper I, our posterior distributions of $K$ give $\langle K \rangle \sim 0.55$ (and hence $\beta\approx 0$).  Figure~\ref{fig:betmK} highlights the importance of $K$ in modeling elliptical galaxies.

Therefore, if we accept the existence of $M_\star/L$ gradients (see references in the Introduction), then we must conclude that previous spherical Jeans dynamical modeling which ignored gradients is likely to be biased towards larger $\beta$:  i.e., towards finding more radial anisotropy.  

Figure~2 of \cite{Ber18} shows why $\beta$ and $K$ are expected to be anti-correlated. For the observed \siglos\ profile and light distribution, the \siglos\ profile cannot in general be fitted by an isotropic velocity dispersion and a constant $M_\star/L$. The observed \siglos\ profile can then be tried to be fitted by a radially varying $\beta$ or a non-zero $K$. For example, a \siglos\ profile that is rising towards the center can be realized by, either a relatively higher $\beta$ along with a relatively less steep mass profile, or a relatively lower $\beta$ along with a relatively steeper mass profile. As a result, there is a degeneracy between $\beta$ and $K$ that controls the steepness of the mass profile for the given light profile. Comparison of cases (b) and (c) shows that this degeneracy is in part broken by the observed \siglos\ profile in some cases such as NGC~4486 (Figure~\ref{fig:VP4486}). Only a $K>0$ can have the required strong effect in the central region to fit the rising \siglos\ profile well. On the other hand, for a case like NGC~4753 whose \siglos\ profile is not rising towards the center, a $K>0$ does not improve the fit, but a radially varying $\beta$ is required (c.f.\ Table~\ref{tab:chisq}).

This $K$-$\beta$ degeneracy is a modern version of the classical mass-anisotropy degeneracy first discussed in \cite{BM82}. The increased precision and spatial-resolution that are now available let us study the interplay between the {\em profiles} of $\beta$, $M_\star/L$, and the DM.

\subsection{Degeneracy between $f_{\rm DM}$ and $M_\star/L$ gradient}
Our analysis has shown that the $K$-$\beta$ degeneracy persists even when DM is included.  Therefore, we now show what our results imply for the DM distribution.

As Figure~1 of \cite{Ber18} shows, when $K$ is increased (i.e. the $M_\star/L$ gradient is stronger), then the stellar mass distribution becomes more centrally concentrated, so the DM mass within $R_{\rm e}$ must increase to fit the observed line-of-sight velocity dispersions outside the central region.  In addition, Figures 17 and 18 of Paper~I show that the average $M_\star/L$ within $R_{\rm e}$ decreases.

\begin{figure*} 
  \vspace{1em}
   \hspace{0em}
  \includegraphics[angle=-90,origin=c,scale=0.38]{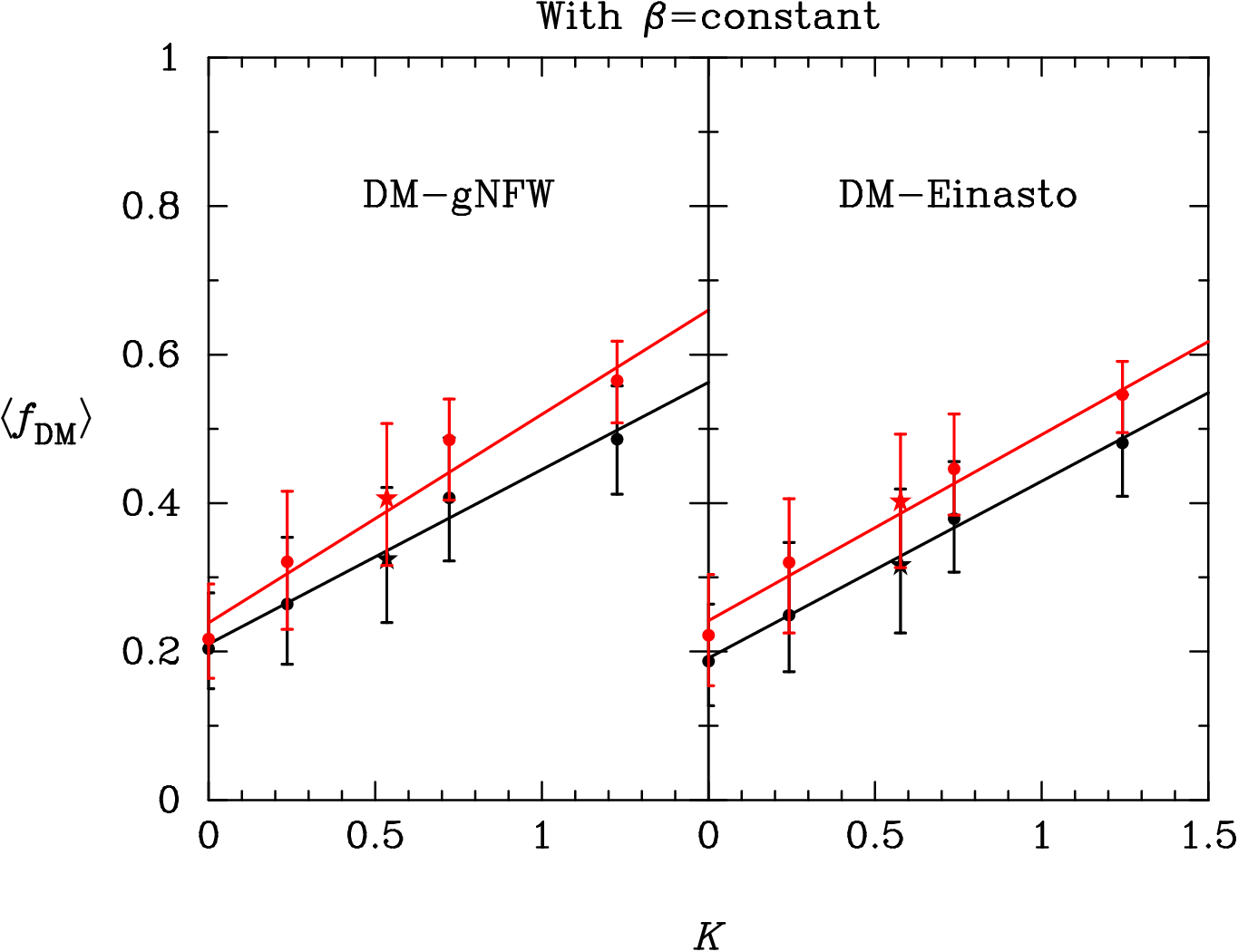}
  \hspace{1em}
  \includegraphics[angle=-90,origin=c,scale=0.38]{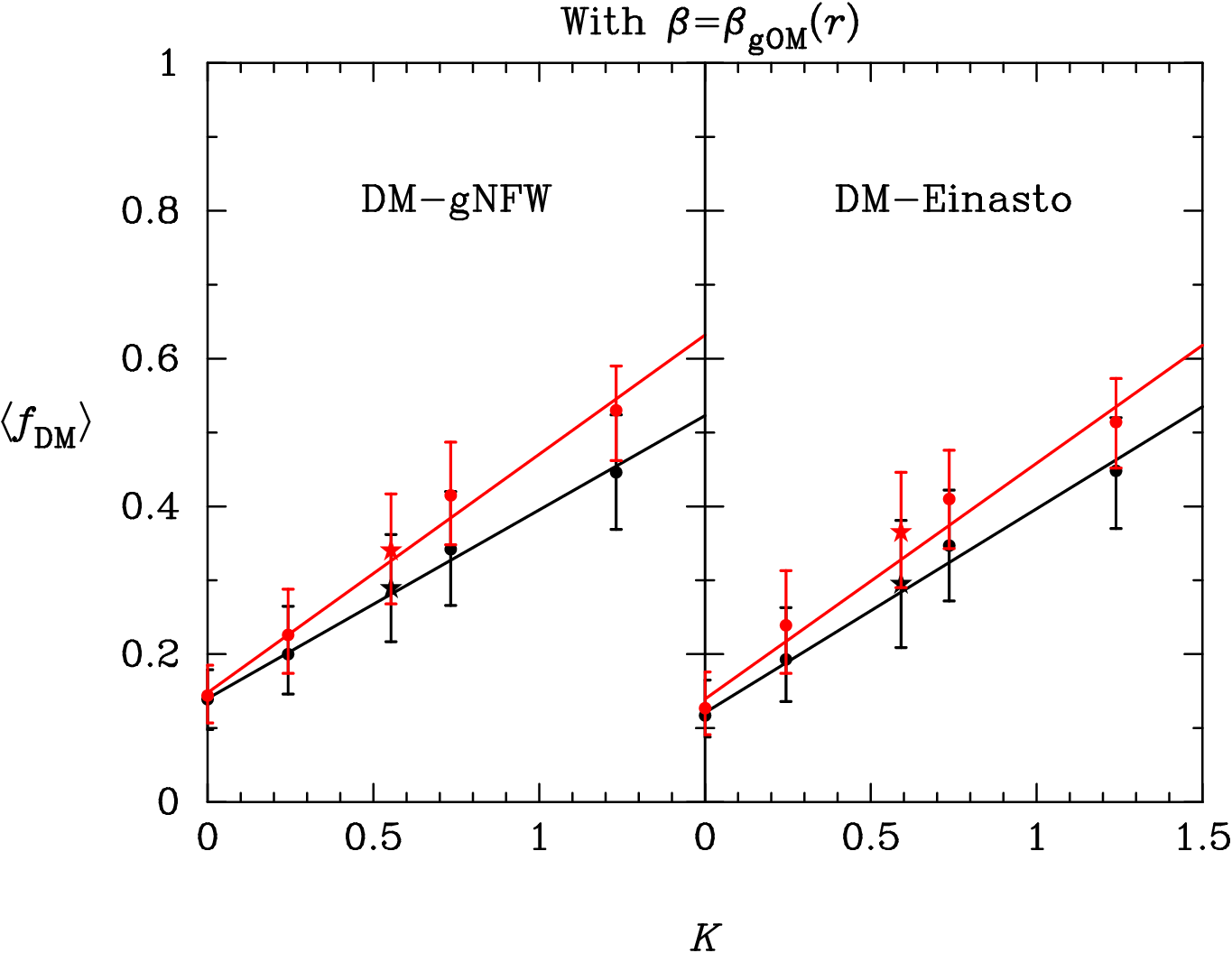}

  \vspace{-2em}
\caption{Median DM fractions within a sphere of $r=R_{\rm e}$ with respect to $K$ with constant anisotropies (left) and the gOM anisotropy model (right). The results are shown for both the gNFW DM model and the Einasto DM model. Black points and lines represent all 24 pure-bulges while red ones represent the 16 slow rotators.  \label{fig:fdmK}}
\end{figure*}

Figure~\ref{fig:fdmK} exhibits this expected scaling of the DM fraction $f_{\rm DM}$ (within a sphere of $r=R_{\rm e}$) with $K$. Together, the gNFW and Einasto results imply
\begin{equation}
  \langle f_{\rm DM} \rangle \approx a_f + b_f K,
\end{equation}
with:

 $a_f = 0.20 \pm 0.05$, $b_f=0.22\pm 0.07$ (all), and 

 $a_f = 0.25 \pm 0.05$, $b_f=0.26\pm 0.07$ (SRs)\\
when $\beta={\rm constant}$; 

 $a_f = 0.14 \pm 0.03$, $b_f=0.26\pm 0.07$ (all), and

 $a_f = 0.16 \pm 0.03$, $b_f=0.31\pm 0.06$ (SRs)\\
when  $\beta=\beta_{\rm gOM}(r)$.

Marginalizing over $K$ yields:

$\langle f_{\rm DM} \rangle = 0.32\pm 0.09$ (all) and
$0.40\pm 0.09$ (SRs)\\
when $\beta={\rm constant}$;

$\langle f_{\rm DM} \rangle = 0.30\pm 0.08$ (all) and
$0.35\pm 0.08$ (SRs) \\
when $\beta=\beta_{\rm gOM}(r)$.
These DM fractions are larger than the value $f_{\rm DM}\sim 0.13$ returned by JAM modeling of these galaxies \citep{Cap13}.  However, if we, like they, set $K=0$, then our analysis also returns $f_{\rm DM}\sim 0.13$ for all pure-bulges.  Table~\ref{tab:MsL} gives the values of $f_{\rm DM}$, $\Upsilon_{\star{\rm e}}$ (the average value of $M_\star/L$ within the projected $R_{\rm e}$) as well as $\Upsilon_{\star 0}$ (the constant value at $R>0.4 R_{\rm e}$) for cases (a-d) with the gNFW DM model specified in Table~\ref{tab:chisq}.

\subsection{Implication for MOND}
The increased DM fraction in the optical regions when $K \neq 0$ under the $\Lambda$CDM paradigm must imply an important modification to the MOND IF compared with the case for $K=0$. This means that $M_\star/L$ gradient is a crucial factor in studying the RAR using elliptical galaxies. The reader is referred to \citet{CBS18b} for a detailed discussion of the RAR based on extensive modeling results including those considered here.   

\subsection{Correlation of $\beta$ with velocity dispersions}
\begin{figure*} 
  \vspace{2em}
   \hspace{1em}
  \includegraphics[angle=-90,origin=c,scale=0.37]{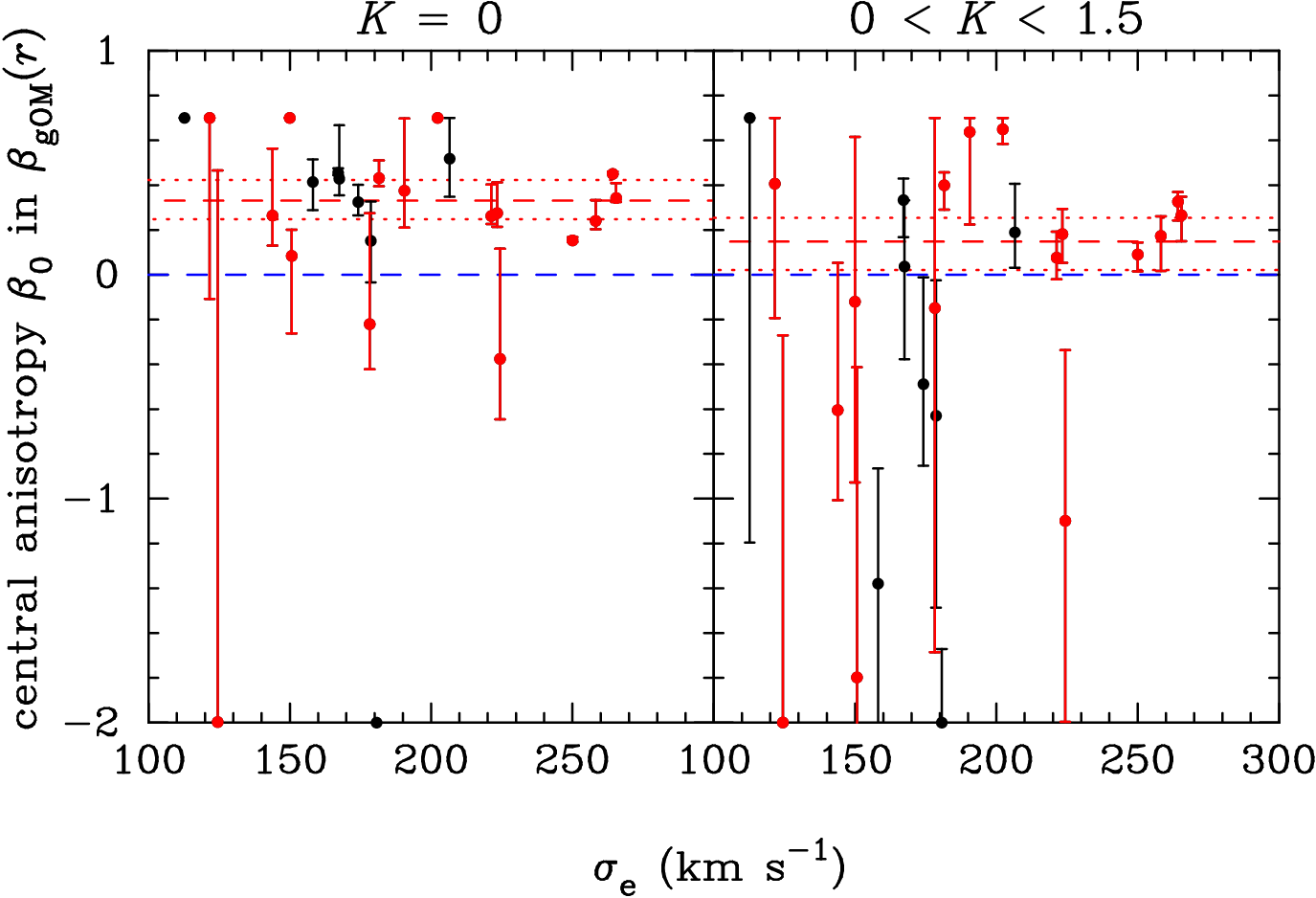}
   \hspace{2em}
  \includegraphics[angle=-90,origin=c,scale=0.37]{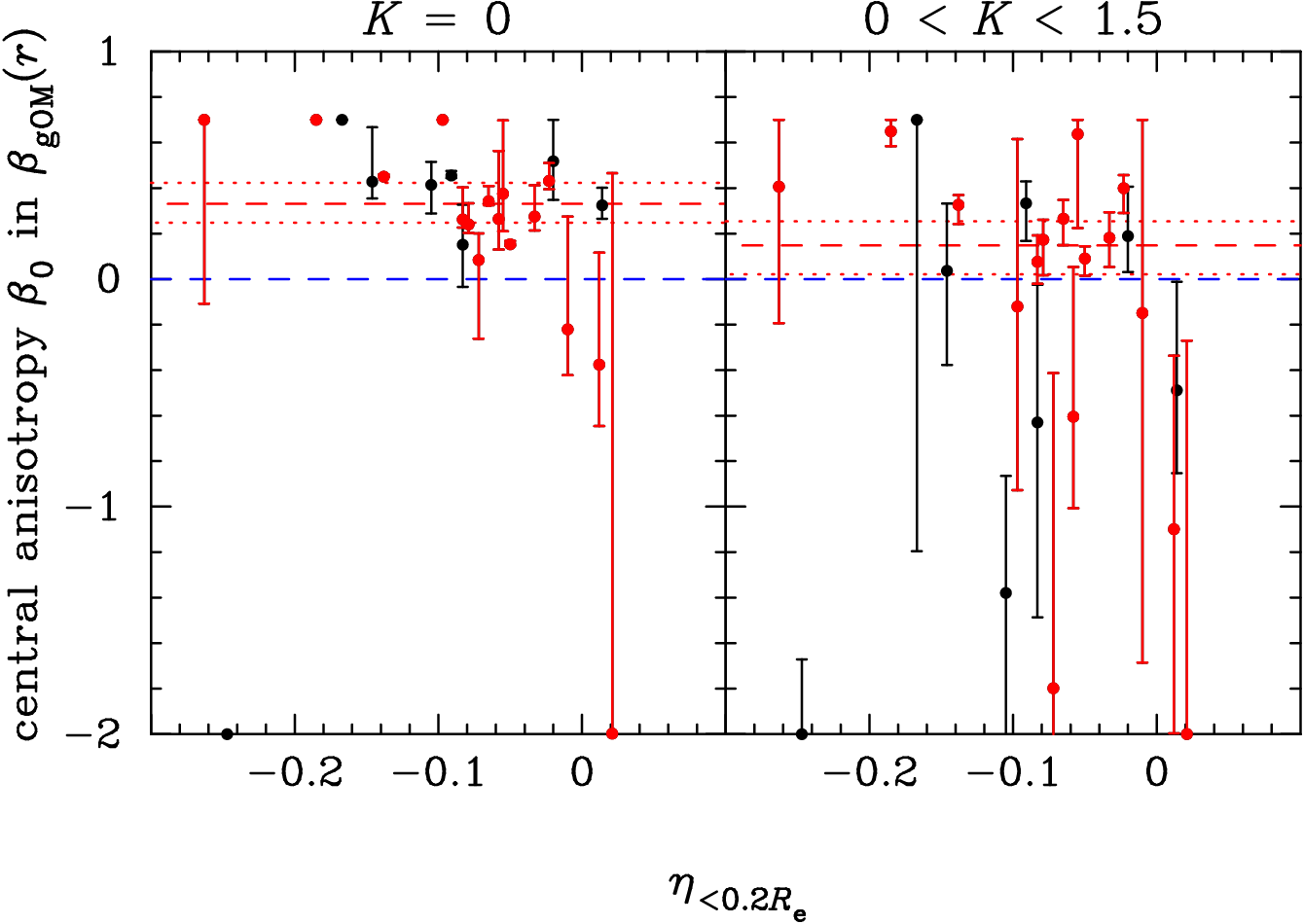}

   \vspace{-2em}
\caption{(Left) Central anisotropies with respect to the effective velocity dispersion $\sigma_{\rm e}$ of the 24 {\atl} pure-bulges for the cases of (b) and (d) of Figure~\ref{fig:bet4gNFW}. For case (d), the higher-$\sigma_{\rm e}$ galaxies and the lower-$\sigma_{\rm e}$ galaxies show a dichotomy. (Right) Central anisotropies with respect to $\eta_{<0.2 R_{\rm e}}$. \label{fig:bet0gNFW}}
\end{figure*}

\begin{figure*} 
  \vspace{1em}
  
  \hspace{-5em}
\includegraphics[scale=0.65]{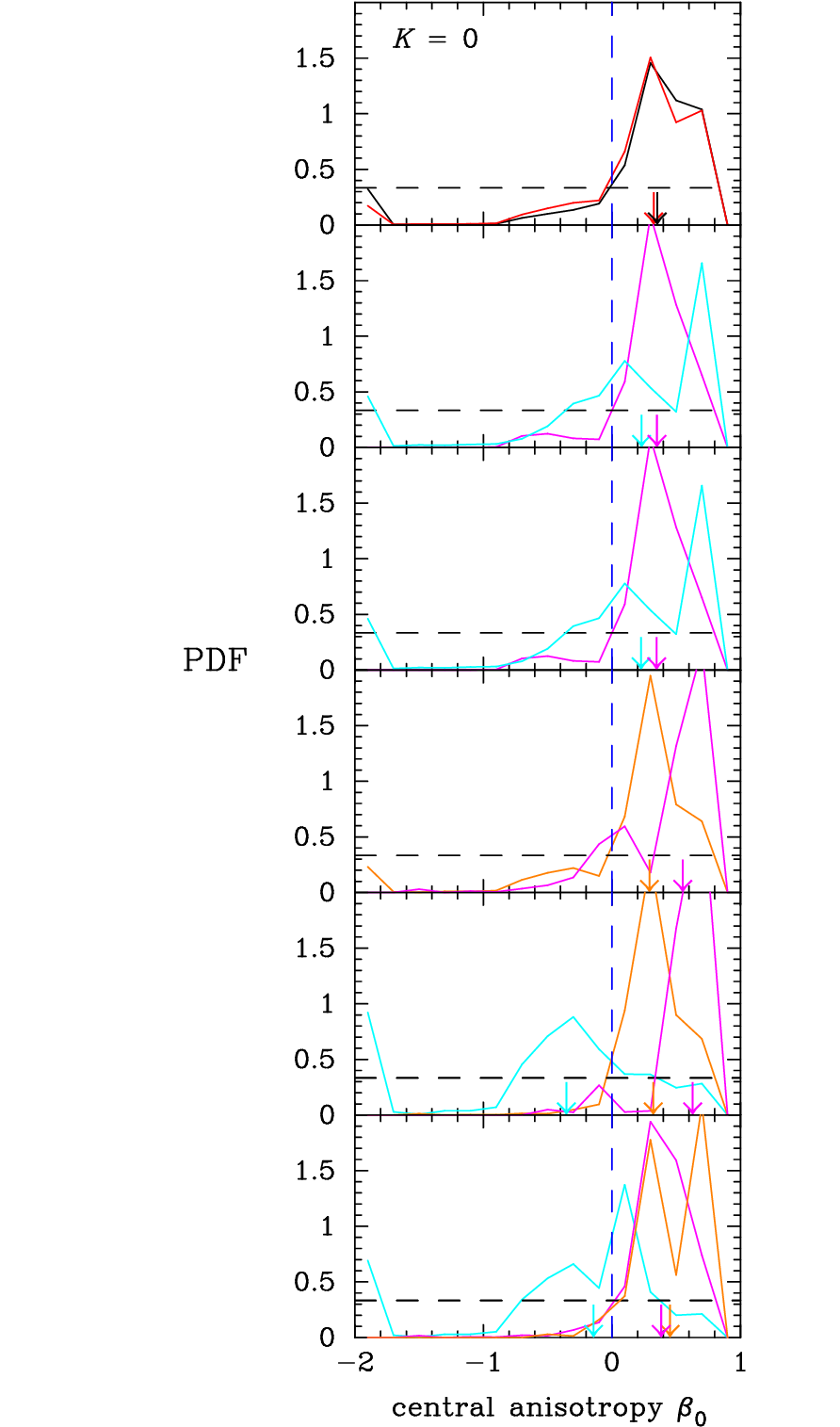}\includegraphics[scale=0.65]{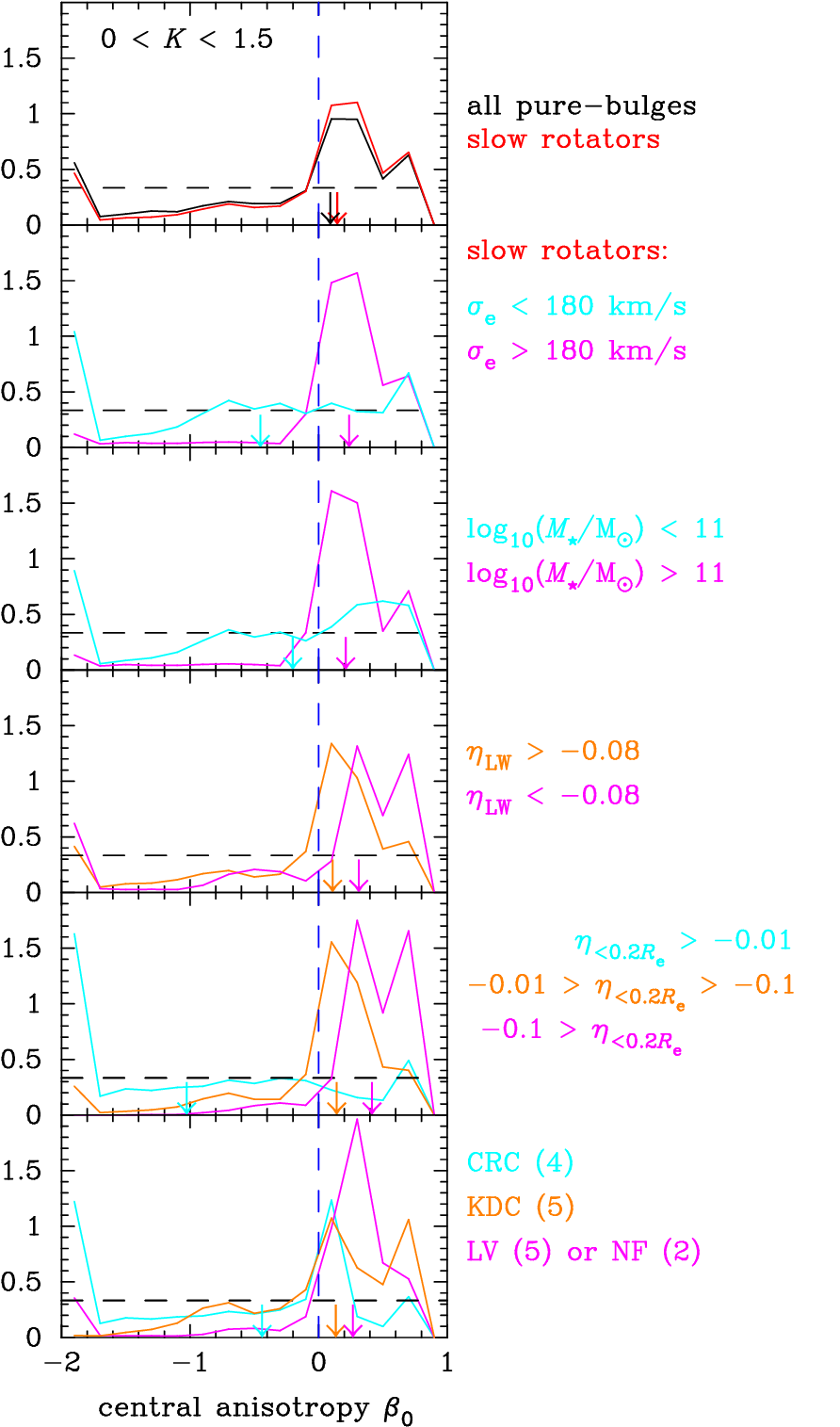}\caption{Same as Figure~\ref{fig:pdfbetm}, but for the central anisotropy $\beta_0$. \label{fig:pdfbet0}}
\end{figure*}

For the most general case, (d), Figures~\ref{fig:bet4gNFW} and~\ref{fig:pdfbetm} hint that higher-$\sigma_{\rm e}$ ($\sigma_{\rm e}>180$~{\kms}) and lower-$\sigma_{\rm e}$ ($\sigma_{\rm e}<180$~{\kms}) galaxies may have different anisotropies, as would be expected if their formation histories are different \citep{Xu17,Li18}. If we consider the central anisotropy parameter $\beta_0$, then the dichotomy appears clearer, as shown in Figures~\ref{fig:bet0gNFW} and~\ref{fig:pdfbet0}.
There appears a similar dichotomy between the higher stellar mass ($M_\star > 10^{11} M_\odot)$ sub-sample and the lower  stellar mass ($M_\star < 10^{11} M_\odot)$ sub-sample. The higher-$\sigma_{\rm e}$ (or $M_\star$) galaxies are likely to be radially biased while the lower-$\sigma_{\rm e}$ (or $M_\star$) galaxies have a significant probability of (or, even prefers to) being tangentially biased.

\subsection{Correlation of $\beta$ with the slope of {\siglos}}
\begin{figure*} 

  \vspace{-3em}
  \hspace{8em}  \includegraphics[angle=-90,origin=c,scale=0.55]{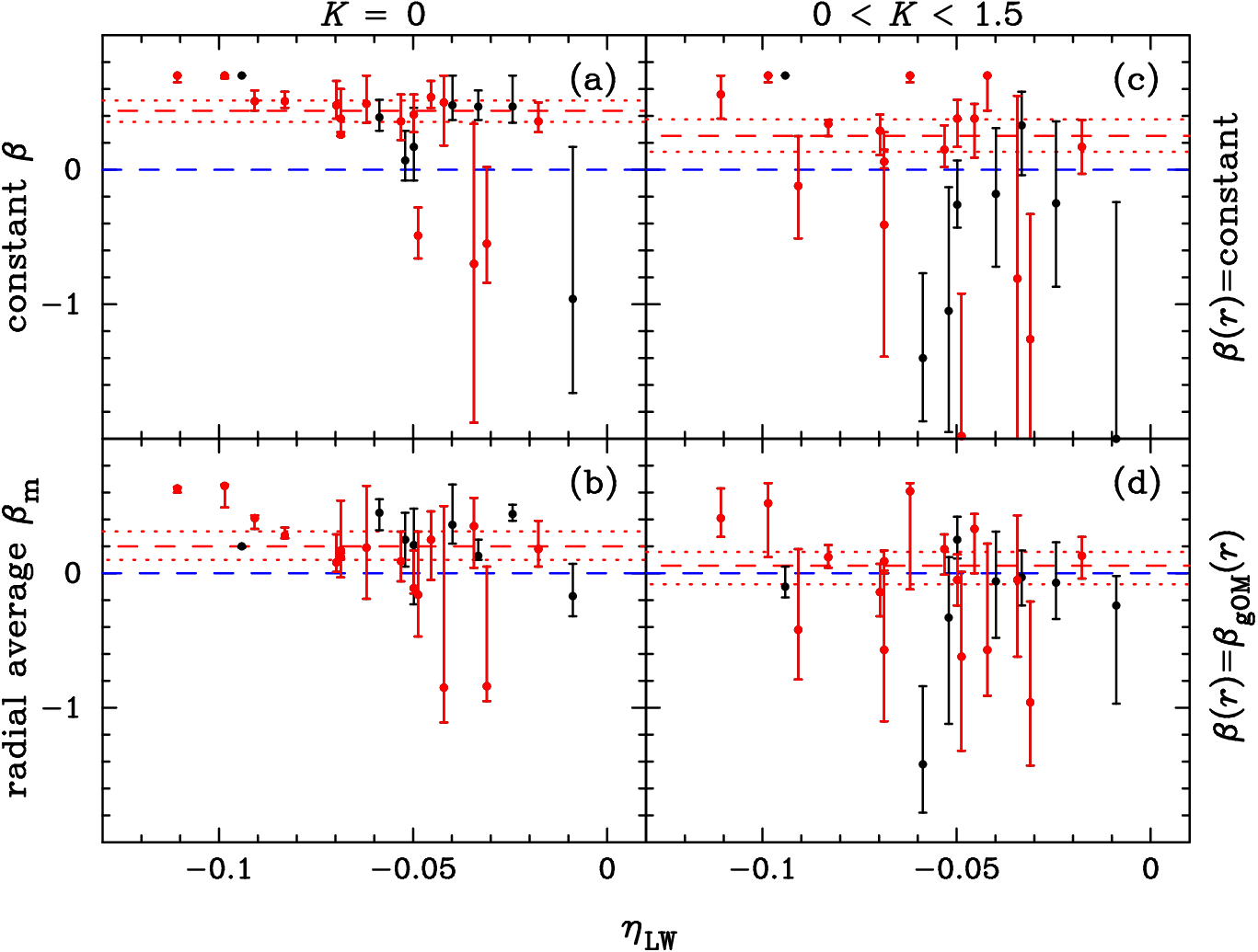}

  \vspace{-3em}
  \hspace{8em}  \includegraphics[angle=-90,origin=c,scale=0.55]{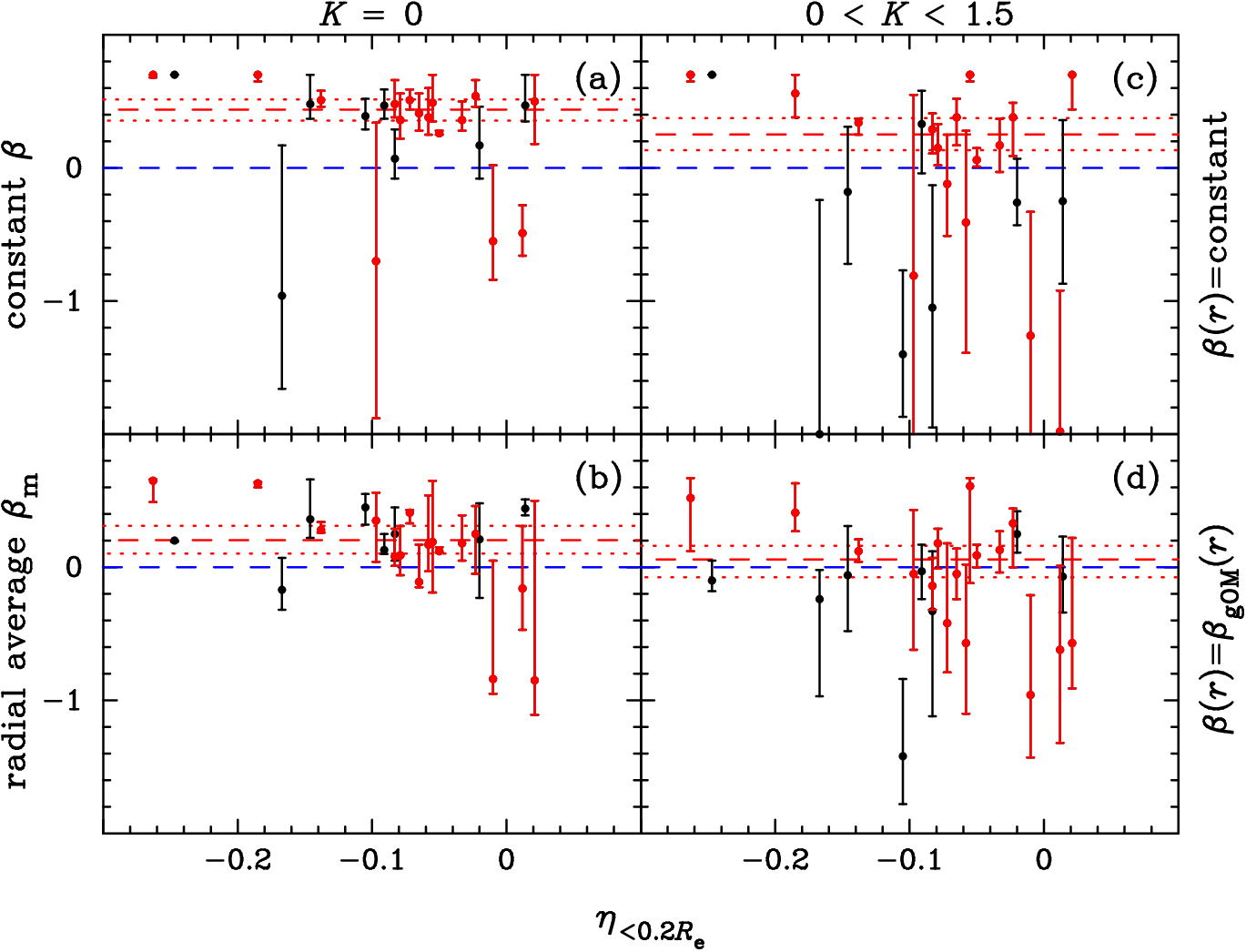}

  \vspace{-4em}
\caption{Same as Figure~\ref{fig:bet4gNFW}, but with respect to the slopes (see Table~\ref{tab:chisq}) $\eta_{\rm LW}$ (upper) and  $\eta_{<0.2 R_{\rm e}}$ (lower). \label{fig:bet4eta}}
\end{figure*}

Traditionally, the line-of-sight velocity dispersions are light-weighted within a projected radius $R$ and this light-weighted velocity dispersion, denoted by $\langle\sigma_{\rm los}\rangle(R)$, is empirically approximated by a power-law relation with $R$, i.e.\  $\langle\sigma_{\rm los}\rangle(R)\propto R^{\eta_{\rm LW}}$ \citep{Jor95}. We calculate $\eta_{\rm LW}$ using two values, $\sigma_{\rm e}$ at $R=R_{\rm e}$ and $\sigma_{\rm e/8}$ at $R=R_{\rm e}/8$ taken from \citet{Cap13,Cap13b} as reproduced in Table~\ref{tab:chisq}. The calculated values of $\eta_{\rm LW}$ are given in the Table: they exhibit a large galaxy-to-galaxy scatter in the range $-0.13\la \eta_{\rm LW} \la 0$ with a median of $\langle\eta_{\rm LW}\rangle \approx -0.06$ \citep{Cap06}. The upper part of Figure~\ref{fig:bet4eta} shows the fitted anisotropies with respect to $\eta_{\rm LW}$ and the fourth panels from the top of Figure~\ref{fig:pdfbetm} show the PDFs for sub-samples divided by $\eta_{\rm LW}$.  The upper left-hand panels of Figure~\ref{fig:bet4eta} show that galaxies with steeper velocity dispersions ($\eta_{\rm LW}>-0.08$) all appear to be radially biased when $K=0$.  The fourth left-hand panel of Figure~\ref{fig:pdfbetm} shows the shift in the value of $\beta_{\rm m}$ between the steeper and the shallower sub-samples for $K=0$. This too can be understood in the context of the degeneracy between $\beta$ and $K$ (c.f.\ Figure~2 of \citet{Ber18}).  Indeed, when we allow for $0<K<1.5$ there is no longer a noticeable dependence on $\eta_{\rm LW}$ (see the upper right-hand panels of Figure~\ref{fig:bet4eta} or the fourth right-hand panel of Figure~\ref{fig:pdfbetm}).  

The likely more interesting and relevant quantity is the true (as opposed to the light-weighted) slope of the {\siglos} profile in the central region. This slope exhibits a greater diversity among the observed {\siglos} profiles of SSBGs (see, e.g., Figure~6 of Paper I).  Whereas most elliptical galaxies exhibit negative slopes (i.e.\ velocity dispersions usually decline with $R$), some observed and simulated elliptical galaxies exhibit flat or inverted {\siglos} profiles near the center (i.e. they increase with $R$).  These are referred to as central dips or depressions in the literature (see, e.g., \citealt{Naab14}).

We consider the slope $\eta_{<0.2 R_{\rm e}}$ within $R < 0.2 R_{\rm e}$ defined by $\sigma_{\rm los}(R)\propto R^{\eta_{<0.2 R_{\rm e}}}$. Table~\ref{tab:chisq} gives the measured values of $\eta_{<0.2 R_{\rm e}}$ based on the {\siglos} profiles shown in Figure~6 of Paper~I. The lower part of Figure~\ref{fig:bet4eta} exhibits the fitted anisotropies with respect to the central slope $\eta_{<0.2 R_{\rm e}}$.  In all four cases, the three SSBGs with steepest negative slopes ($\eta_{<0.2 R_{\rm e}}<-0.1$) are all significantly radially biased. Furthermore, for the cases of the varying anisotropy (cases~(b) and (d)), three SSBGs with positive or flat slopes ($\eta_{<0.2 R_{\rm e}}>-0.01$: we consider this relaxed cut considering the measurement uncertainties) are all significantly tangentially biased. The fifth panels from the top in Figure~\ref{fig:pdfbetm} show the PDFs. When the SSBGs are split into three bins of $\eta_{<0.2 R_{\rm e}}$ a systematic trend of $\beta_{\rm m}$ with $\eta_{<0.2 R_{\rm e}}$ is evident. When non-zero $K$ is allowed, galaxies with positive or flat central slopes are even more tangentially biased. 

The right-hand panel of Figure~\ref{fig:bet0gNFW} exhibits the fitted central anisotropies which are expected to be more directly related to the slope $\eta_{<0.2 R_{\rm e}}$. The fifth panels from the top in Figure~\ref{fig:pdfbet0} show the PDFs of $\beta_0$ for three bins of $\eta_{<0.2 R_{\rm e}}$. Indeed, the systematic trend of $\beta_0$ with $\eta_{<0.2 R_{\rm e}}$ is stronger than $\beta_{\rm m}$. For the realistic case of marginalizing $K$ over $0<K<1.5$ there is a clear dichotomy between the steep slope ($\eta_{<0.2 R_{\rm e}}<-0.1$) sample and the flat or inverted slope ($\eta_{<0.2 R_{\rm e}}>-0.01$) sample. The latter is tangentially biased with $\langle\beta_0\rangle\approx -1.0$ while the former is radially biased with $\langle\beta_0\rangle\approx 0.4$.

\subsection{Correlation with kinemetry}
Our results indicate that central features of the velocity dispersions \citep{Kra11,Naab14} in SRs are closely related to the anisotropies of the orbital distributions. To understand why we compare $\eta_{<0.2 R_{\rm e}}$ with kinematic features of the observed line-of-sight velocity dispersions as measured and classified by \citet{Kra11} based on the so-called kinemetry analysis. \citet{Kra11} classified all 260 {\atl} ETGs into six groups based on kinematic features. According to their classification ETGs are broadly divided into regular rotators (RRs) and non-regular rotators (NRRs). Most of the NRRs are SRs based on the angular momentum parameter $\lambda_{\rm e}$ of \citet{Ems11}. The NRRs exhibit specific features such as kinematically distinct cores (KDCs), counter-rotating cores (CRCs), and low-level (rotational) velocities (LVs). KDCs mean cores whose rotation (although rotation itself is small for SRs) axes shift abruptly (more than $30\deg$) from the surrounding regions. In the transition regions there are no detectable rotations. When the shift is of the order of $180\deg$, they are called CRCs (thus CRCs are extreme cases of KDCs). LVs refer to low-level rotation velocities throughout the observed regions. Out of our selected sample of 16 SSBGs, we have four SRs with CRCs, five with KDCs, five with LVs and two with no features (NFs) as given in Table~\ref{tab:chisq}. 

\begin{figure}
\begin{center}
\setlength{\unitlength}{1cm}
\begin{picture}(10,11)(0,0)
\put(-0.6,-0.5){\includegraphics{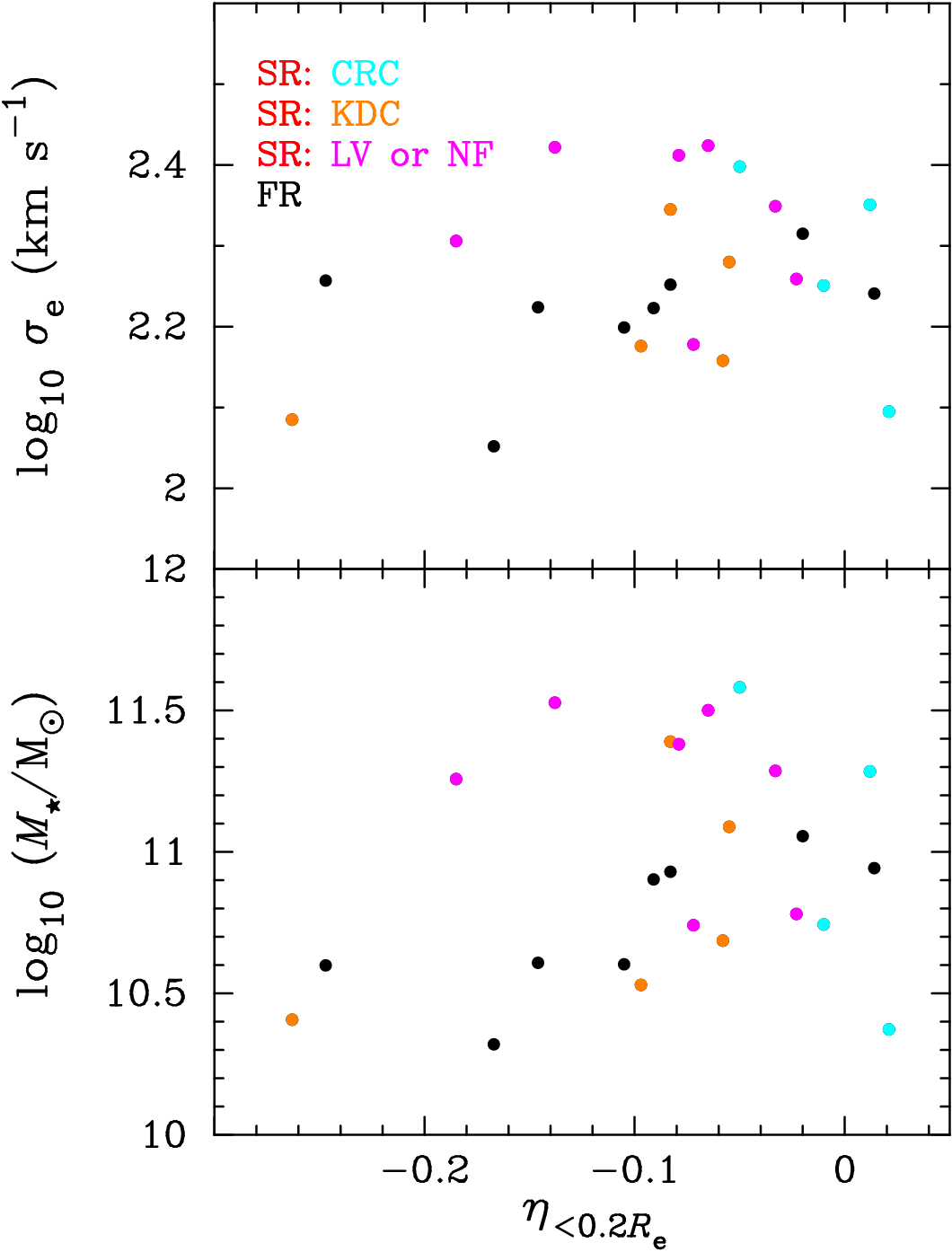}}
\end{picture}
\caption{Kinematic features (see Table~\ref{tab:chisq}) of pure-bulges with respect to $\eta_{<0.2R_{\rm e}}$ (the line-of-sight velocity dispersion slope in the central region: see Table~\ref{tab:chisq}), $\sigma_{\rm e}$ (the light-weighted line-of-sight velocity dispersions within $R_{\rm e}$ taken from \citet{Cap13}), and $M_{\star}$ (our fitted stellar mass for the case of (d): see Table~\ref{tab:chisq}). \label{fig:etacen}}
\end{center}
\end{figure}

Figure~\ref{fig:etacen} exhibits all 24 {\atl} pure-bulges with respect to $\eta_{<0.2 R_{\rm e}}$, $\sigma_{\rm e}$, and $M_\star$, coding kinematic features with different colors. SRs with CRCs have shallower or inverted slopes compared with other kinematic classes of SRs. In particular, all three SRs with positive/flat slopes ($\eta_{<0.2 R_{\rm e}}>-0.01$) are CRCs, but not all CRCs have positive/flat slopes.  Figure~\ref{fig:etacen} shows the well-known fact that SRs are more massive and have higher velocity dispersions than FRs. It also shows that kinematic features of SRs (i.e.\ CRCs, KDCs, and LVs) do not have preferences for $\sigma_{\rm e}$ or $M_\star$. The only apparent correlation is that CRCs are biased towards the higher side of  $\eta_{<0.2 R_{\rm e}}$ compared with KDCs and LVs/NFs. Then, the correlation of anisotropies with $\eta_{<0.2 R_{\rm e}}$ shown in Figure~\ref{fig:pdfbetm} implies that CRCs are likely to be more tangentially biased in the central regions. 

The bottom right-hand panel of Figure~\ref{fig:pdfbetm} shows the PDFs of anisotropies for three kinematic classes of SRs. After marginalizing $K$ over $0< K < 1.5$, the three classes show distinct features in $\beta$. SRs with CRCs are more likely to be tangentially biased but exhibit dual possibilities, i.e.\ tangential and isotropic (or mildly radial). SRs with KDCs are on average isotropic with individual possibilities of radial or tangential anisotropies. LVs/NFs are likely to be radially biased or isotropic over the IFS probed regions ($\la R_{\rm e}$), but are clearly radially biased at the center as $\beta_0 > 0$ without exception as shown in the bottom right-hand panel of Figure~\ref{fig:pdfbet0}.

These results from our modeling of SSBGs provide reasonable dynamical explanations for the observed kinematic features of SRs. The fact that three out of four CRCs have positive/flat central slopes of {\siglos} profiles (``central depressions'') with tangential anisotropies while one has a (``normal'') negative central slope with isotropic or mildly radially biased velocity dispersions (c.f.\ Tables \ref{tab:chisq} and \ref{tab:beta}) hints at two distinct origins for CRCs. CRCs with positive/flat slopes are probably dissipationally formed cores with tangentially dominated orbits which naturally give rise to rising or flat {\siglos} profiles in the central region. CRCs with declining {\siglos} profiles require another explanation. In Section~\ref{sec:sim} we use recent cosmological hydrodynamic and merger simulations to discuss dynamical origins of CRCs.

LVs (note that NFs are similar to LVs but with more rotations) are systems with no detectable net rotation over the regions $\la R_{\rm e}$. Systems with predominantly random (chaotic) orbits will have isotropic orbits, but our finding that LVs are more likely to be radially biased particularly in the centers mean that the orbits have not been randomized and consist more of infalling orbits. KDCs can be viewed as intermediate systems between CRCs and LVs, hence both possibilities of radial and tangential biases are equally likely.

\section{Comparison with Cosmological Simulations} \label{sec:sim}

Recent cosmological hydrodynamic simulations of galaxy formation and evolution and galaxy merger simulations make specific predictions regarding kinematic and dynamic properties in the optical regions of galaxies, and connect these properties with formation and evolution histories. Although there exist a number of state-of-the-art cosmological simulations, here we consider only those simulations that investigate kinematic features and velocity dispersion anisotropies of elliptical galaxies. A key aspect of our dynamical modeling is to allow for radial gradients in $M_\star/L$ in the region $R<0.4 R_{\rm e}$ based on a host of recent reports (see Section~\ref{sec:intro}). However, no cosmological simulations have allowed for such a possibility so far. Hence, comparison of our modeling results with currently available simulations is somewhat limited. As we discuss below, we find qualitative agreement but also some quantitative tension. Nevertheless, we find that such simulations are quite useful in interpreting our modeling results.

Two simulations are most relevant to the present discussion. One is the cosmological zoom-in simulation by \citet{Oser10} and the other is the Illustris simulation \citep{Vog14a,Vog14b,Gen14}. Based on the cosmological zoom-in simulation \citet{Naab14} classified 44 zoomed-in ETGs into six groups with distinctive formation paths and present-day photometric and kinematic properties. \citet{Rott14} then investigate the stellar orbits of these simulated galaxies and provide predicted anisotropy profiles $\beta(r)$ for all six groups. The zoom-in simulation provides finest details of the kinematic and dynamic properties of simulated galaxies at present. One caveat of the zoom-in simulation is that their galaxies are not naturally representative of galaxies formed from a large volume simulation. This caveat is complemented by the Illustris simulation which is a simulation of a periodic box of $106.5$~Mpc on a side. The Illustris simulation can provide distributions of averaged anisotropies with respect to galaxy properties such as the light-weighted velocity dispersion $\sigma_{\rm e}$ (or $\sigma_{\rm e/2}$) and the in-situ stellar mass fraction as well as (less detailed) anisotropy profiles. Such analyses have been carried out by \citet{Wu14}, \citet{Xu17}, and  \citet{Li18}. 

We first discuss the overall properties of the anisotropies of SRs and then discuss the classes grouped by kinematic features. The simulations predict that the anisotropies of all (i.e.\ in-situ {\it plus} accreted that are observed) stellar motions for {\it all} SRs over the radial range $<R_{\rm e}$ (corresponding to our probed regions) are radially biased $\beta(r)>0$. See Figures~18 and 19 of \citet{Rott14} in which their classes C, E and F are SRs, and also Figure~11 of \citet{Wu14} in which larger in-situ fractions can reduce $\beta(r)$ but still remain radially biased. These results are in qualitative agreement with the corresponding results of our modeling case~(b) for which $M_\star/L$ is assumed to be constant but anisotropy is allowed to vary radially: see Figures~\ref{fig:bet4gNFW} and \ref{fig:betgOMcML}. For case~(b), the median radially averaged anisotropy is $\approx 0.2$ which is in good agreement with both simulations. However, real galaxies allow tangential anisotropies -- at least two galaxies in our sample (NGC~0661 and NGC~1289) prefer tangential biases (Figure~\ref{fig:betgOMcML}). For a large sample of simulated ETGs \citet{Li18} find a correlation of $\beta_{\rm m}$ with $\sigma_{\rm e/2}$ so that lower-$\sigma_{\rm e/2}$ galaxies may have $\beta_{\rm m}<0$. However, they do not distinguish fast and slow rotators so it is not clear whether they include any SRs having $\beta_{\rm m}<0$.

What is more striking is that when $M_\star/L$ gradients are allowed ($K>0$), the median anisotropy for SRs gets close to zero with both radial and tangential biases occurring with nearly equal probabilities. The only simulations which currently incorporate IMF driven $M_\star/L$ gradients are those of \cite{Bar18}.  Applying our analysis to their simulations is beyond the scope of this work, but is ongoing. What follows is a discussion of the origin of this isotropy based on previously available simulation results.

When our selected 16 {\atl} SSBGs are grouped by kinematic features as identified by \citet{Kra11}, they are divided into three groups, i.e.\ 4 CRCs, 5 KDCs, and 7 LVs/NFs. Three CRCs have positive/flat central slopes (``central depressions'') of line-of-sight velocity dispersions while one does not as shown in Figure~\ref{fig:etacen}. SRs with central depressions belong to Class~C defined by \citet{Naab14}. Orbit analyses by \citet{Rott14} show that Class~C galaxies have the lowest anisotropies among SRs (their Figure~19) although still radially biased. This agrees qualitatively with our anisotropy results shown in Figures~\ref{fig:pdfbetm} and \ref{fig:pdfbet0}. However, our results show that SRs with central depressions are likely to be tangentially biased or isotropic without exception, although the rest are radially biased or isotropic when $K=0$ is assumed as in simulations. For the realistic case of allowing $0<K<1.5$, SRs with central depressions are more likely to be tangentially biased. According to \citet{Naab14} Class~C galaxies have undergone late gas-rich major mergers. Central depressions are thought to originate from ``stars that have formed from gas driven to the center of the galaxy during the merger, a process well studied in isolated binary mergers \citep{BH96}.'' Therefore, a dissipationally formed core that is kinematically decoupled from the main body is likely to have more tangentially biased orbits. These galaxies also have relatively higher in-situ fractions that are also consistent with the general trends seen in the Illustris simulation \citep{Wu14,Xu17}. Interestingly, the simulated galaxies with central depressions from the cosmological zoom-in simulations \citep{Naab14} do not exhibit CRCs while our selected three {\atl} SRs with central depressions exhibit CRCs without exception. Note, however, that galaxy merger simulations have reproduced central depressions exhibiting CRCs (e.g.\ \citealt{BQ90,Jess07,Tsat15}). The remaining one SR with a CRC from our sample does not exhibit a central depression. This galaxy might be consistent with Class~E galaxies by \citet{Naab14} which have undergone gas-poor major mergers. These comparisons suggest that SRs with CRCs have been formed through recent major mergers with or without gas dissipation that determines the feature of central depression. 

LVs are the galaxies with no kinematic features with no significant rotations (NFs are similar to LVs but with some angular momenta). These galaxies may be consistent with Class~F and Class~E of \citet{Naab14} that have undergone only dry (minor and/or major) mergers. These galaxies have low in-situ (and thus high accreted) fractions of stars. Both hydrodynamics \citep{Rott14,Wu14,Xu17} and merger \citep{Hilz12} simulations generically predict that those galaxies have radially biased orbits qualitatively consistent with our modeling results. Note here that those simulations have not considered possibilities of $M_\star/L$ gradients. Our results for 7 LVs/NFs with $K=0$ give $\langle\beta_{\rm m}\rangle\approx 0.4$ with a broad possible range of $-0.2\la \beta_{\rm m} \la 0.7$ while Figure~19 of \citet{Rott14} gives $0.1<\beta_{\rm e/2}<0.4$ for 14 Class~F/E simulated galaxies. While the predicted median is similar to our results, the simulation predicts too narrow a range of anisotropies. When $M_\star/L$ gradients are allowed (the right-hand side of our Figure~\ref{fig:pdfbetm}), the predicted median is lower ($\langle\beta_{\rm m}\rangle \approx 0.2$) but the possible range is similar. Interestingly, our results for the central anisotropy $\beta_0$ (Figure~\ref{fig:pdfbet0}) give $\beta_0>0$ for all LVs/NFs regardless of the assumption on $M_\star/L$ gradients, while that is not the case for other kinds of SRs for $K\ne 0$ (the right-hand side of Figure~\ref{fig:pdfbet0}). This is consistent with the picture that LVs/NFs do not keep dissipationally-formed central components as would be lost from dry mergers. Perhaps, this is not a surprising result because LVs/NFs do not contain any distinct kinematic feature in the central regions by their kinematic definition. Radially biased orbits in the central regions imply that infalling orbits from accreted stars are dominating.    

KDCs are weaker versions of CRCs (or CRCs are extreme versions of KDCs). KDCs may not exactly belong to any of classes of SRs identified by \citet{Naab14}. They may be assigned to an intermediate class between Class~C and Class~E/F. Note that \citet{Naab14} used just 44 simulated galaxies which may not include the real variety of elliptical galaxies. Our results show that their median properties are close to isotropic, but both tangential and radial biases are occurring individually.

\section{Summary, Discussion and Conclusions} \label{sec:conc}

We have investigated the velocity dispersion anisotropy profiles of 24 {\atl} pure-bulge galaxies paying particular attention to 16 nearly spherical, slowly-rotating, pure-bulge galaxies. These SSBGs constitute an extreme subset of ETGs (recall that most ETGs are now known to exhibit some rotation as revealed by IFS studies \citep{Ems11,Kra11}). Therefore, our anisotropy results cannot be representative for general ETGs. Nevertheless, they reveal key aspects of the dynamical structure of SRs and provide unique constraints on the astrophysics of galaxy formation and evolution.

Empirically, SRs tend to be more massive than FRs among ETGs. In the standard $\Lambda$CDM model, SRs are the end products of the hierarchical process of galaxy formation and evolution. Therefore, their present-day orbital structure not only reveals their current dynamical state but also contains information about their dynamical (thus formation and evolution) history.  Early in-situ star formation, growth by gas-rich or poor mergers, late merger-driven in-situ formation, the effects of feedback from supernovae and AGN, etc., are all expected to influence the current dynamics of SRs.

We have estimated velocity dispersion anisotropies for these galaxies under four different assumptions (Table~\ref{tab:chisq}) about the anisotropy radial profile and the $M_\star/L$ radial gradient, for each of four models of DM halos or MOND IFs. For the simplest case -- a constant $M_\star/L$ and radially constant anisotropy -- the observed line-of-sight velocity dispersions cannot be well fitted for $\sim 40$\% of the galaxies, and the fitted anisotropies are clearly radially biased for most of the modeled galaxies (case (a) in Table~\ref{tab:chisq} and Figure~\ref{fig:bet4gNFW}). However, for the most realistic case -- $M_\star/L$ gradient strength $K$ is marginalized over $0<K<1.5$ and anisotropy is radially varying with the flexible form of the gOM model (Equation~(\ref{eq:gOM})) -- all but one of the galaxies can be successfully modeled (and even for this one case, the line-of-sight velocity dispersions can be fitted reasonably) and the median anisotropy is close to zero; i.e.\ isotropy is preferred (case (d) in Table~\ref{tab:chisq} and Figure~\ref{fig:bet4gNFW}). This holds for all four models of DM halos or MOND IFs. Here $M_\star/L$ gradients play a key role, as shown by the systematic trend of the median anisotropy $\langle\beta_{\rm m}\rangle$ with $K$ (Figure~\ref{fig:betmK}).  This can be understood as follows. If $M_\star/L$ increases towards the center, but it is modeled assuming there is no gradient, then isotropic velocity dispersions in the central regions would give line-of-sight velocity dispersions {\siglos} that are flatter than observed (see, e.g., Figure~6 of Paper I). Radial anisotropies must then be invoked to match the observed steepening, but our results suggest that they are artifacts of ignoring $M_\star/L$ gradients (also see Figure~2 of \citealt{Ber18}).

Under the $\Lambda$CDM paradigm $M_\star/L$ gradients also have important consequences for DM distributions. If $M_\star/L$ is larger in the central regions ($\la 0.4R_{\rm e}$), then the DM contribution to the total mass distribution must be enhanced (Figure~\ref{fig:fdmK}) while the average $M_\star/L$ gets lowered to match the observed {\siglos} profiles on scales of order $R_e$ and larger.  Thus, for SRs, $M_\star/L$ gradient, velocity dispersion anisotropy and DM distribution are closely related.

Given the isotropy of SRs in the median sense we have also investigated possible correlations of the fitted anisotropies with various other properties of galaxies. We find that the anisotropies are well correlated with the slopes of the line-of-sight velocity dispersions in the central regions ($<0.2 R_{\rm e}$) denoted by $\eta_{<0.2R_{\rm e}}$ (Figures~\ref{fig:bet4eta} and \ref{fig:bet0gNFW}). SRs with steeper $\eta_{<0.2R_{\rm e}}$ are more radially anisotropic. On the other hand, SRs with flat or inverted slopes ($\eta_{<0.2R_{\rm e}} > -0.01$) are likely to be tangentially biased (Figures~\ref{fig:pdfbetm} and \ref{fig:pdfbet0}).

We also notice that $\eta_{<0.2R_{\rm e}}$ correlates with kinemetric features of SRs identified by \citet{Kra11}. SRs with central depressions in the line-of-sight velocity dispersions (in the sense of having inverted or flat slopes $\eta_{<0.2R_{\rm e}} > -0.01$) have CRCs without exception (see Figure~\ref{fig:etacen} and Table~\ref{tab:chisq}). However, there exist SRs with CRCs (one case out of our four SRs with CRCs) that do not have central depressions (see \citet{Kra11}). SRs having KDCs, LVs or NFs have steep slopes $\eta_{<0.2R_{\rm e}} < -0.02$ (Figure~\ref{fig:etacen}). SRs grouped by three kinematic features of CRCs, KDCs and LVs/NFs have systematically different anisotropies (Figures~\ref{fig:pdfbetm} and \ref{fig:pdfbet0}): CRCs are tangentially biased or isotropic (or mildly radial) while LVs/NFs are radially biased or isotropic (or mildly tangential). KDCs are close to isotropic in the median sense. This systematic trend is most pronounced in the most general and realistic modeling case (case~(d) of Table~\ref{tab:chisq} and Figure~\ref{fig:bet4gNFW}) and is at odds with the predictions by currently available simulations. Two main shortcomings of the existing simulations are the incomplete treatment of feedback from supernovae and AGN, and the use of a fixed stellar IMF.  The latter means that simulations underestimate variations in $M_\star/L$ across the galaxy population, as well radial gradients within individual galaxies.

Although currently available cosmological simulations cannot be directly compared with the anisotropies obtained here for SRs, they \citep{Naab14,Rott14,Wu14,Xu17,Li18} can be used to interpret our anisotropy results and make connections with formation and evolution histories of SRs. SRs with CRCs have shallowest $\eta_{<0.2R_{\rm e}}$ (Figure~\ref{fig:etacen}) and often inverted slopes ($\eta_{<0.2R_{\rm e}}\ga 0$). Our results show that objects with positive slopes tend to be tangentially biased (Figures~\ref{fig:pdfbetm} and \ref{fig:pdfbet0}. Tangentially biased orbits are consistent with scenarios in which the galaxies have undergone major gas-rich mergers that resulted in forming cores that are decoupled from the main bodies, as would be realized for Class~C galaxies by \citet{Naab14}. However, Class~C simulated galaxies are not counter-rotating. Moreover, when \citet{Rott14} looked into velocity dispersion anisotropy profiles of these galaxies, they obtained only radially-biased orbits for the regions $\la R_{\rm e}$ (although Class~C galaxies had the lowest anisotropies among SRs). These discrepancies between our results for SRs with $\eta_{<0.2R_{\rm e}}\ga 0$ and the Class~C galaxies of \citet{Naab14} are likely to be consequences of the shortcomings of the existing simulations as pointed out above. CRCs without central depressions are not likely to be tangentially biased (although our small sample includes just one CRC without central depression). Such galaxies may have been formed through gas-poor major mergers as would be realized for Class~E simulated galaxies by \citet{Naab14} which indeed includes a counter-rotating case.

We find that SRs with LVs/NFs, which constitute the largest fraction of SRs \citep{Kra11}, are likely to be radially biased. Our results are qualitatively consistent with outputs from major and/or minor gas-poor (dry) mergers (Class E/F galaxies of \citet{Naab14}) which have lowest angular momenta. However, our results allow a broader possibility including mild tangential biases (Figures~\ref{fig:pdfbetm} and \ref{fig:pdfbet0}) while the simulations produced only radial biases that are strongest among all classes of ETGs. SRs with KDCs are intermediate in their kinematic features (as KDCs are weaker versions of CRCs) and our results show that their anisotropies are also intermediate. This means that SRs with KDCs are most likely to be isotropic (or mildly radially biased) in the median sense with a broad possible range. 

We have explicitly allowed for $M_\star/L$ gradients for $R<0.4 R_{\rm e}$ in Jeans dynamical analyses of nearly spherical pure-bulge galaxies. We find that $M_\star/L$ gradients have significant impacts on the inference of the velocity dispersion anisotropies. When $M_\star/L$ gradients are marginalized over a reasonable range, the median anisotropy of SRs is zero which is not the case in previous dynamical modeling results without $M_\star/L$ gradients. Furthermore, SRs with different kinematic features have systematically different anisotropies. Thus, the isotropy in the median sense does not represent a dynamical property such as chaotic orbits, but emerges as a coincidence arising from various classes. These results cannot yet be reproduced by existing cosmological simulations. Our investigations call for the need to consider $M_\star/L$ gradients in dynamical modeling and cosmological simulations. 

Our present work suffers from two caveats. One is the assumption of spherical symmetry and the other is small sample size.  While triaxial models would be better representations of pure-bulge galaxies, the fact that most of our selected nearly spherical pure-bulge {\atl} galaxies were successfully modeled under the spherical symmetry assumption suggests that the spherical symmetry assumption is not too unrealistic. The issue of sample size can be addressed by applying our analysis to galaxies in the MaNGA \citep{Bun15} survey, which will provide an order of magnitude more galaxies like those studied here. This will allow us to apply even stricter criteria when selecting galaxies so that we can test the effects of varying selection criteria under the spherical symmetry assumption.  We intend to do this in the near future.  As galaxy formation simulations which include gradients become available, we will use them to test our Jeans equation-based analysis.  The first simulations to incorporate IMF driven $M_\star/L$ gradients have only just been completed \citep{Bar18}.  We expect to report on the results of applying our analysis to their simulations in the near future.

In conclusion, from a range of MC models of 24 nearly spherical pure-bulge {\atl} galaxies, of which 16 are kinematic SRs, we have obtained the following results:

\begin{enumerate}
\item If the stellar mass-to-light ratio ($M_\star/L$) is assumed to be constant within a galaxy, then one is likely to conclude that SRs have radially-biased orbits, with a median spherical anisotropy of $\langle\beta_{\rm m}\rangle \approx 0.2$.  This is in good agreement with the literature results.

\item However, if $M_\star/L$ is allowed to have a radial gradient for $R<0.4 R_{\rm e}$ and the strength of this gradient is marginalized over the currently allowed range, SRs are consistent with being isotropic.

\item If $M_\star/L$ gradients are allowed, then the DM contribution to the total mass distribution in the central region ($<R_{\rm e}$) of SRs is $\langle f_{\rm DM}\rangle \sim 0.35$.  This is about twice as large as when gradients are ignored.  As a result, SRs may in fact provide interesting probes of MOND.  
  
\item The median isotropy of SRs appears to be a coincidence arising from a diversity of anisotropies for different kinematic SR sub-classes rather than a typical dynamical property of SRs.

\item The diverse anisotropies of SRs have much to do with the diverse slopes ($\eta_{<0.2R_{\rm e}}$) of the line-of-sight velocity dispersions in the central regions. SRs with very steep slopes ($\eta_{<0.2R_{\rm e}}<-0.1$) are radially biased while SRs with flat or inverted slopes ($\eta_{<0.2R_{\rm e}}>-0.01$ representing central depressions) are tangentially biased.

\item Three out of four SRs with CRCs exhibit central depressions and thus are tangentially biased. One SR with a CRC that does not exhibit central depression is not tangentially biased. SRs with CRCs may have been formed through major mergers and the amount of gas involved (i.e.\ whether gas-rich or gas-poor) may have influenced the presence or the lack of central depression.

\item SRs with LVs or NFs are likely to be radially biased. They may have been formed through gas-poor major and/or minor mergers.
  
\item SRs with KDCs are intermediate between SRs with CRCs and SRs with LVs. Their velocity dispersions are close to the isotropy in the median sense.
  
\end{enumerate}

\acknowledgments

We thank the referee for a number of useful comments that helped us improve the manuscript significantly, and Thorsten Naab for helpful comments regarding his simulations. KHC acknowledges support by Basic Science Research Program through the National Research Foundation of Korea (NRF) funded by the Ministry of Education (NRF-2016R1D1A1B03935804). MB thanks NSF AST/1816330 for support.

\appendix

\section{Tables of Fitted Quantities}

\begin{deluxetable*}{ccccccccccc}[b]
\tablecaption{Fitted anisotropies of various cases with the gNFW DM model. \label{tab:beta}}
\renewcommand\thetable{A1}
\tabletypesize{\scriptsize}
\tablewidth{0pt}
\tablehead{
  \colhead{galaxy} &  \colhead{}  &   \colhead{(a)}  &  \colhead{} &  \multicolumn{2}{c}{(b)} &  \colhead{}  &  \colhead{(c)}  &   \colhead{} &    \multicolumn{2}{c}{(d)} \\
   \cline{3-3} \cline{5-6} \cline{8-8} \cline{10-11}
 \colhead{} &  \colhead{} &   \colhead{ $\beta$}  &  \colhead{} & \colhead{$\beta_{\rm m}$}   & \colhead{$\beta_{0}$}  &  \colhead{} & \colhead{$\beta$} &  \colhead{} & \colhead{$\beta_{\rm m}$} & \colhead{$\beta_{0}$}
}
\startdata
  NGC 0661 &   &  $-0.55^{+0.56}_{-0.29}$ &   & $-0.84^{+0.89}_{-0.11} $ & $-0.22^{+0.50}_{-0.20}$  &   & $-1.26^{+0.93}_{-0.74} $  &   & $-0.96^{+0.75}_{-0.47} $   & $-0.15^{+0.85}_{-1.54}$   \\
  NGC 1289 &   &   $0.50^{+0.20}_{-0.32}$ &    & $-0.85^{+1.35}_{-0.26} $  & $-2.00^{+2.46}_{-0.00}$   &    & $0.70^{+0.00}_{-0.25} $  &   & $-0.57^{+0.79}_{-0.34} $   & $-2.00^{+1.73}_{-0.00}$   \\
  NGC 2695 &   &   $0.70^{+0.00}_{-0.00}$ &    & $0.20^{+0.01}_{-0.00} $  & $-2.00^{+0.00}_{-0.00}$   &    & $0.70^{+0.00}_{-0.00} $  &   & $-0.10^{+0.15}_{-0.08} $   & $-2.00^{+0.33}_{-0.00}$   \\
  NGC 3182 &   &   $-0.96^{+1.12}_{-0.70}$ &    & $-0.17^{+0.24}_{-0.16} $  & $ 0.70^{+0.00}_{-0.00}$  &    & $-2.00^{+1.76}_{-0.00} $ &    & $-0.24^{+0.22}_{-0.73} $    & $0.70^{+0.00}_{-1.90}$   \\
  NGC 3193 &   &  $0.07^{+0.22}_{-0.15} $ &     & $0.25^{+0.20}_{-0.20} $  & $ 0.15^{+0.18}_{-0.18}$  &    & $-1.05^{+0.92}_{-0.90} $  &   & $-0.33^{+0.45}_{-0.79} $   & $-0.63^{+0.60}_{-0.86}$    \\
  NGC 3607 &   &  $0.17^{+0.29}_{-0.24} $ &     & $0.21^{+0.26}_{-0.44} $  & $ 0.52^{+0.18}_{-0.17}$  &     & $-0.26^{+0.33}_{-0.17} $  &   & $0.25^{+0.18}_{-0.14} $   & $ 0.19^{+0.22}_{-0.16}$   \\
  NGC 4261 &   &   $0.41^{+0.16}_{-0.12} $ &     & $-0.11^{+0.28}_{-0.04} $  & $ 0.34^{+0.07}_{-0.02}$  &     & $0.38^{+0.14}_{-0.21} $  &   & $-0.05^{+0.19}_{-0.19} $   & $ 0.26^{+0.08}_{-0.11}$   \\
  NGC 4365 &   &  $0.48^{+0.19}_{-0.10} $ &     & $0.08^{+0.22}_{-0.07} $  & $ 0.26^{+0.14}_{-0.03}$  &     & $0.29^{+0.12}_{-0.18} $  &   & $-0.14^{+0.21}_{-0.18} $  & $ 0.08^{+0.12}_{-0.10}$    \\
  NGC 4374 &   &   $0.36^{+0.20}_{-0.13} $ &     & $0.09^{+0.22}_{-0.14} $  & $ 0.24^{+0.09}_{-0.04}$  &     & $0.15^{+0.18}_{-0.13} $  &   & $0.18^{+0.11}_{-0.19} $   & $ 0.17^{+0.09}_{-0.16}$   \\
  NGC 4406 &   &  $0.49^{+0.21}_{-0.14} $ &     & $0.19^{+0.45}_{-0.38} $  & $ 0.38^{+0.32}_{-0.17}$   &    & $0.70^{+0.00}_{-0.05} $  &   & $0.61^{+0.07}_{-0.73} $  & $ 0.64^{+0.06}_{-0.41}$    \\
  NGC 4459 &   &  $0.39^{+0.13}_{-0.10} $  &    & $0.45^{+0.10}_{-0.13} $  & $ 0.41^{+0.10}_{-0.13}$  &     & $-1.40^{+0.63}_{-0.47} $ &    & $-1.42^{+0.58}_{-0.36} $   & $-1.38^{+0.51}_{-0.62}$   \\
  NGC 4472 &   &   $0.26^{+0.02}_{-0.01} $ &     & $0.12^{+0.03}_{-0.01} $  & $ 0.15^{+0.02}_{-0.01}$   &    & $0.06^{+0.09}_{-0.06} $ &    & $0.09^{+0.08}_{-0.09} $  & $ 0.09^{+0.05}_{-0.08}$    \\
  NGC 4486 &   &   $0.51^{+0.07}_{-0.06}  $ &    & $0.28^{+0.06}_{-0.02} $  & $ 0.45^{+0.01}_{-0.00}$   &    & $0.34^{+0.03}_{-0.08} $ &    & $0.12^{+0.09}_{-0.08} $   & $ 0.33^{+0.04}_{-0.08}$   \\
  NGC 4636 &   &  $0.54^{+0.12}_{-0.08} $  &    & $0.25^{+0.22}_{-0.30} $  & $ 0.43^{+0.08}_{-0.04}$   &    & $0.38^{+0.11}_{-0.29} $ &    & $0.33^{+0.12}_{-0.33} $   & $ 0.40^{+0.06}_{-0.11}$   \\
  NGC 4753 &   &   $0.47^{+0.22}_{-0.12} $ &     & $0.44^{+0.07}_{-0.06} $   & $ 0.32^{+0.08}_{-0.06}$   &    & $-0.25^{+0.61}_{-0.63} $ &    & $-0.07^{+0.29}_{-0.28} $  & $-0.49^{+0.48}_{-0.36}$   \\
  NGC 5322 &   &  $-0.49^{+0.21}_{-0.17} $ &     & $-0.16^{+0.48}_{-0.30} $  & $-0.38^{+0.49}_{-0.27}$   &    & $-1.98^{+1.06}_{-0.02} $ &    & $-0.62^{+0.64}_{-0.70} $   & $-1.10^{+0.76}_{-0.90}$   \\
  NGC 5481 &   &  $0.70^{+0.00}_{-0.02} $ &     & $0.65^{+0.00}_{-0.16} $  & $ 0.70^{+0.00}_{-0.81}$   &    & $0.70^{+0.00}_{-0.05} $ &    & $0.52^{+0.15}_{-0.40} $   & $ 0.41^{+0.29}_{-0.60}$   \\
  NGC 5485 &   &  $0.47^{+0.12}_{-0.10} $ &     & $0.13^{+0.12}_{-0.03} $  & $ 0.46^{+0.02}_{-0.01}$   &    & $0.33^{+0.26}_{-0.36} $ &    & $-0.03^{+0.19}_{-0.21} $   & $ 0.33^{+0.10}_{-0.17}$   \\
  NGC 5557 &   &  $0.70^{+0.00}_{-0.05} $ &     & $0.63^{+0.01}_{-0.03} $  & $ 0.70^{+0.00}_{-0.00}$   &    & $0.56^{+0.14}_{-0.18}  $ &    & $0.41^{+0.22}_{-0.14} $   & $ 0.65^{+0.05}_{-0.06}$   \\
  NGC 5631 &   &   $-0.70^{+1.04}_{-1.18} $ &     & $0.35^{+0.21}_{-0.31} $  & $ 0.70^{+0.00}_{-0.00}$  &    & $-0.81^{+1.36}_{-1.19} $ &    & $-0.05^{+0.48}_{-0.57} $    & $-0.12^{+0.74}_{-0.81}$   \\
  NGC 5831 &   &  $0.38^{+0.22}_{-0.14} $ &     & $0.17^{+0.37}_{-0.21} $  & $ 0.26^{+0.30}_{-0.13}$   &    & $-0.41^{+0.69}_{-0.98} $ &    & $-0.57^{+0.58}_{-0.53} $   & $-0.61^{+0.66}_{-0.40}$   \\
  NGC 5846 &   &  $0.36^{+0.14}_{-0.09} $ &     & $0.18^{+0.21}_{-0.12} $  & $ 0.28^{+0.14}_{-0.06}$   &    & $0.17^{+0.20}_{-0.20} $ &    & $0.13^{+0.14}_{-0.17}  $   & $ 0.18^{+0.11}_{-0.13}$   \\
  NGC 5869 &   &  $0.48^{+0.22}_{-0.11} $ &     & $0.36^{+0.29}_{-0.15} $  & $ 0.43^{+0.24}_{-0.07}$  &     & $-0.18^{+0.49}_{-0.54} $ &    & $-0.06^{+0.36}_{-0.42} $   & $ 0.04^{+0.30}_{-0.41}$   \\
  NGC 6703 &   &  $0.51^{+0.08}_{-0.06} $ &     & $0.41^{+0.02}_{-0.08} $  & $ 0.08^{+0.12}_{-0.35}$   &    & $-0.12^{+0.37}_{-0.39} $ &    & $-0.42^{+0.60}_{-0.37} $   & $-1.80^{+1.38}_{-0.20}$   \\
\enddata
\tablecomments{Fitted anisotropy values for the four different cases of Table~\ref{tab:chisq}. For cases (b) and (d), the parameter $\beta_{\rm m}$ and $\beta_0$, respectively, refer to the radially averaged value and the central value of the fitted gOM model (Equation~(\ref{eq:gOM})).}
\end{deluxetable*}

\begin{longrotatetable}
  \begin{deluxetable*}{ccccccccccccccc}   
\tablecaption{Fitted $M_\star/L$ and $f_{\rm DM}(r=R_{\rm e})$ of various cases with the gNFW DM model. \label{tab:MsL}}
\renewcommand\thetable{A2}
\tablewidth{0pt}
\tabletypesize{\scriptsize}
\tablehead{
  \colhead{galaxy} &  \colhead{} &   \multicolumn{2}{c}{(a)}  & \colhead{} &  \multicolumn{2}{c}{(b)}  & \colhead{} &   \multicolumn{3}{c}{(c)}  & \colhead{} &   \multicolumn{3}{c}{(d)} \\
   \cline{3-4} \cline{6-7} \cline{9-11} \cline{13-15}
 \colhead{} &  \colhead{} &  \colhead{ $f_{\rm DM}$} & \colhead{$\log_{10}\Upsilon_{\star 0}$} & \colhead{}  &  \colhead{ $f_{\rm DM}$} & \colhead{$\log_{10}\Upsilon_{\star 0}$}  & \colhead{} & \colhead{ $f_{\rm DM}$} & \colhead{$\log_{10}\Upsilon_{\star 0}$} & \colhead{$\log_{10}\Upsilon_{\star{\rm e}}$} & \colhead{} & \colhead{$f_{\rm DM}$} & \colhead{$\log_{10}\Upsilon_{\star 0}$} & \colhead{$\log_{10}\Upsilon_{\star{\rm e}}$}
}
\startdata
NGC 0661 &   & $  0.103^{+  0.205}_{ -0.063}$ & $  0.911^{+  0.024}_{ -0.111}$ & \colhead{} & $  0.066^{+  0.205}_{ -0.05}$ & $  0.944^{+  0.018}_{ -0.118}$ & \colhead{} & $  0.237^{+  0.146}_{ -0.129}$ & $  0.792^{+  0.074}_{ -0.12}$ & $  0.821^{+  0.058}_{ -0.083}$ & \colhead{} & $  0.211^{+  0.177}_{ -0.154}$ & $  0.770^{+  0.132}_{ -0.126}$ & $  0.818^{+  0.101}_{ -0.103} $ \\
NGC 1289 &   & $  0.407^{+  0.151}_{ -0.074}$ & $  0.413^{+  0.056}_{ -0.126}$ & \colhead{} & $  0.208^{+  0.229}_{ -0.103}$ & $  0.515^{+  0.019}_{ -0.118}$ & \colhead{} & $  0.551^{+  0.155}_{ -0.23}$ & $  0.251^{+  0.156}_{ -0.31}$ & $  0.274^{+  0.13}_{ -0.262}$ & \colhead{} & $  0.431^{+  0.19}_{ -0.186}$ & $  0.290^{+  0.138}_{ -0.17}$ & $  0.331^{+  0.113}_{ -0.136} $ \\
NGC 2695 &   & $  0.035^{+  0.028}_{ -0.021}$ & $  0.592^{+  0.012}_{ -0.014}$ & \colhead{} & $  0.016^{+  0.009}_{ -0.008}$ & $  0.537^{+  0.006}_{ -0.005}$ & \colhead{} & $  0.025^{+  0.017}_{ -0.015}$ & $  0.530^{+  0.042}_{ -0.019}$ & $  0.599^{+  0.007}_{ -0.01}$ & \colhead{} & $  0.020^{+  0.013}_{ -0.012}$ & $  0.400^{+  0.053}_{ -0.03}$ & $  0.474^{+  0.029}_{ -0.01} $ \\
NGC 3182 &   & $  0.215^{+  0.445}_{ -0.142}$ & $  0.474^{+  0.063}_{ -0.411}$ & \colhead{} & $  0.284^{+  0.287}_{ -0.189}$ & $  0.502^{+  0.054}_{ -0.19}$ & \colhead{} & $  0.412^{+  0.311}_{ -0.227}$ & $  0.252^{+  0.161}_{ -0.294}$ & $  0.336^{+  0.116}_{ -0.292}$ & \colhead{} & $  0.510^{+  0.214}_{ -0.174}$ & $  0.220^{+  0.159}_{ -0.192}$ & $  0.295^{+  0.126}_{ -0.189} $ \\
NGC 3193 &   & $  0.091^{+  0.1}_{ -0.052}$ & $  0.529^{+  0.018}_{ -0.046}$ & \colhead{} & $  0.131^{+  0.229}_{ -0.086}$ & $  0.482^{+  0.044}_{ -0.126}$ & \colhead{} & $  0.099^{+  0.095}_{ -0.058}$ & $  0.456^{+  0.039}_{ -0.059}$ & $  0.489^{+  0.024}_{ -0.04}$ & \colhead{} & $  0.158^{+  0.212}_{ -0.1}$ & $  0.335^{+  0.082}_{ -0.095}$ & $  0.395^{+  0.061}_{ -0.083} $ \\
NGC 3607 &   & $  0.377^{+  0.275}_{ -0.158}$ & $  0.511^{+  0.09}_{ -0.257}$ & \colhead{} & $  0.192^{+  0.321}_{ -0.142}$ & $  0.625^{+  0.057}_{ -0.258}$ & \colhead{} & $  0.346^{+  0.168}_{ -0.095}$ & $  0.491^{+  0.071}_{ -0.132}$ & $  0.519^{+  0.048}_{ -0.122}$ & \colhead{} & $  0.431^{+  0.16}_{ -0.125}$ & $  0.395^{+  0.077}_{ -0.147}$ & $  0.414^{+  0.076}_{ -0.128} $ \\
NGC 4261 &   & $  0.530^{+  0.187}_{ -0.153}$ & $  0.654^{+  0.121}_{ -0.245}$ & \colhead{} & $  0.111^{+  0.274}_{ -0.075}$ & $  0.975^{+  0.015}_{ -0.133}$ & \colhead{} & $  0.700^{+  0.083}_{ -0.044}$ & $  0.393^{+  0.055}_{ -0.235}$ & $  0.470^{+  0.039}_{ -0.23}$ & \colhead{} & $  0.506^{+  0.201}_{ -0.28}$ & $  0.708^{+  0.197}_{ -0.269}$ & $  0.741^{+  0.171}_{ -0.233} $ \\
NGC 4365 &   & $  0.167^{+  0.238}_{ -0.1}$ & $  0.663^{+  0.037}_{ -0.133}$ & \colhead{} & $  0.140^{+  0.198}_{ -0.085}$ & $  0.783^{+  0.018}_{ -0.087}$ & \colhead{} & $  0.449^{+  0.11}_{ -0.132}$ & $  0.465^{+  0.074}_{ -0.113}$ & $  0.519^{+  0.064}_{ -0.075}$ & \colhead{} & $  0.341^{+  0.178}_{ -0.254}$ & $  0.606^{+  0.123}_{ -0.187}$ & $  0.650^{+  0.104}_{ -0.145} $ \\
NGC 4374 &   & $  0.270^{+  0.189}_{ -0.097}$ & $  0.724^{+  0.047}_{ -0.13}$ & \colhead{} & $  0.099^{+  0.17}_{ -0.069}$ & $  0.824^{+  0.031}_{ -0.085}$ & \colhead{} & $  0.290^{+  0.139}_{ -0.06}$ & $  0.688^{+  0.041}_{ -0.097}$ & $  0.697^{+  0.045}_{ -0.072}$ & \colhead{} & $  0.337^{+  0.19}_{ -0.195}$ & $  0.612^{+  0.166}_{ -0.184}$ & $  0.651^{+  0.142}_{ -0.144} $ \\
NGC 4406 &   & $  0.287^{+  0.162}_{ -0.108}$ & $  0.605^{+  0.056}_{ -0.116}$ & \colhead{} & $  0.281^{+  0.419}_{ -0.155}$ & $  0.752^{+  0.082}_{ -0.341}$ & \colhead{} & $  0.744^{+  0.069}_{ -0.164}$ & $  0.145^{+  0.062}_{ -0.191}$ & $  0.213^{+  0.063}_{ -0.154}$ & \colhead{} & $  0.761^{+  0.186}_{ -0.371}$ & $  0.243^{+  0.405}_{ -0.622}$ & $  0.299^{+  0.381}_{ -0.617} $ \\
NGC 4459 &   & $  0.097^{+  0.123}_{ -0.068}$ & $  0.500^{+  0.023}_{ -0.059}$ & \colhead{} & $  0.107^{+  0.154}_{ -0.074}$ & $  0.466^{+  0.043}_{ -0.061}$ & \colhead{} & $  0.068^{+  0.069}_{ -0.042}$ & $  0.328^{+  0.075}_{ -0.029}$ & $  0.414^{+  0.042}_{ -0.016}$ & \colhead{} & $  0.074^{+  0.077}_{ -0.046}$ & $  0.330^{+  0.07}_{ -0.031}$ & $  0.417^{+  0.039}_{ -0.021} $ \\
NGC 4472 &   & $  0.063^{+  0.025}_{ -0.017}$ & $  0.711^{+  0.004}_{ -0.006}$ & \colhead{} & $  0.087^{+  0.058}_{ -0.037}$ & $  0.760^{+  0.004}_{ -0.01}$ & \colhead{} & $  0.270^{+  0.279}_{ -0.196}$ & $  0.580^{+  0.088}_{ -0.202}$ & $  0.623^{+  0.069}_{ -0.146}$ & \colhead{} & $  0.276^{+  0.264}_{ -0.18}$ & $  0.553^{+  0.111}_{ -0.174}$ & $  0.604^{+  0.088}_{ -0.134} $ \\
NGC 4486 &   & $  0.306^{+  0.415}_{ -0.149}$ & $  0.709^{+  0.075}_{ -0.399}$ & \colhead{} & $  0.087^{+  0.249}_{ -0.06}$ & $  0.943^{+  0.014}_{ -0.118}$ & \colhead{} & $  0.626^{+  0.067}_{ -0.074}$ & $  0.423^{+  0.058}_{ -0.135}$ & $  0.491^{+  0.065}_{ -0.107}$ & \colhead{} & $  0.439^{+  0.189}_{ -0.13}$ & $  0.659^{+  0.107}_{ -0.204}$ & $  0.728^{+  0.083}_{ -0.186} $ \\
NGC 4636 &   & $  0.654^{+  0.15}_{ -0.142}$ & $  0.479^{+  0.14}_{ -0.295}$ & \colhead{} & $  0.386^{+  0.298}_{ -0.209}$ & $  0.788^{+  0.152}_{ -0.297}$ & \colhead{} & $  0.852^{+  0.098}_{ -0.303}$ & $  0.116^{+  0.323}_{ -0.499}$ & $  0.135^{+  0.353}_{ -0.461}$ & \colhead{} & $  0.755^{+  0.18}_{ -0.322}$ & $  0.290^{+  0.473}_{ -0.545}$ & $  0.362^{+  0.43}_{ -0.543} $ \\
NGC 4753 &   & $  0.246^{+  0.38}_{ -0.144}$ & $  0.279^{+  0.07}_{ -0.307}$ & \colhead{} & $  0.178^{+  0.131}_{ -0.091}$ & $  0.256^{+  0.038}_{ -0.062}$ & \colhead{} & $  0.205^{+  0.227}_{ -0.109}$ & $  0.233^{+  0.069}_{ -0.104}$ & $  0.312^{+  0.038}_{ -0.108}$ & \colhead{} & $  0.171^{+  0.088}_{ -0.069}$ & $  0.159^{+  0.065}_{ -0.045}$ & $  0.242^{+  0.026}_{ -0.035} $ \\
NGC 5322 &   & $  0.096^{+  0.062}_{ -0.049}$ & $  0.680^{+  0.013}_{ -0.022}$ & \colhead{} & $  0.145^{+  0.248}_{ -0.087}$ & $  0.623^{+  0.042}_{ -0.146}$ & \colhead{} & $  0.115^{+  0.113}_{ -0.06}$ & $  0.625^{+  0.031}_{ -0.086}$ & $  0.646^{+  0.019}_{ -0.059}$ & \colhead{} & $  0.283^{+  0.226}_{ -0.184}$ & $  0.441^{+  0.134}_{ -0.14}$ & $  0.494^{+  0.101}_{ -0.137} $ \\
NGC 5481 &   & $  0.088^{+  0.089}_{ -0.059}$ & $  0.702^{+  0.021}_{ -0.033}$ & \colhead{} & $  0.085^{+  0.078}_{ -0.058}$ & $  0.739^{+  0.019}_{ -0.045}$ & \colhead{} & $  0.142^{+  0.104}_{ -0.098}$ & $  0.605^{+  0.067}_{ -0.059}$ & $  0.675^{+  0.031}_{ -0.038}$ & \colhead{} & $  0.144^{+  0.138}_{ -0.101}$ & $  0.570^{+  0.082}_{ -0.09}$ & $  0.639^{+  0.057}_{ -0.064} $ \\
NGC 5485 &   & $  0.500^{+  0.254}_{ -0.155}$ & $  0.542^{+  0.121}_{ -0.331}$ & \colhead{} & $  0.105^{+  0.2}_{ -0.074}$ & $  0.874^{+  0.021}_{ -0.095}$ & \colhead{} & $  0.516^{+  0.205}_{ -0.156}$ & $  0.478^{+  0.093}_{ -0.321}$ & $  0.527^{+  0.108}_{ -0.345}$ & \colhead{} & $  0.284^{+  0.213}_{ -0.193}$ & $  0.695^{+  0.132}_{ -0.149}$ & $  0.749^{+  0.1}_{ -0.137} $ \\
NGC 5557 &   & $  0.102^{+  0.124}_{ -0.061}$ & $  0.636^{+  0.024}_{ -0.052}$ & \colhead{} & $  0.083^{+  0.07}_{ -0.048}$ & $  0.704^{+  0.028}_{ -0.038}$ & \colhead{} & $  0.178^{+  0.141}_{ -0.089}$ & $  0.534^{+  0.055}_{ -0.079}$ & $  0.601^{+  0.035}_{ -0.053}$ & \colhead{} & $  0.104^{+  0.166}_{ -0.066}$ & $  0.587^{+  0.067}_{ -0.08}$ & $  0.671^{+  0.033}_{ -0.079} $ \\
NGC 5631 &   & $  0.272^{+  0.28}_{ -0.128}$ & $  0.535^{+  0.063}_{ -0.213}$ & \colhead{} & $  0.353^{+  0.229}_{ -0.119}$ & $  0.488^{+  0.099}_{ -0.183}$ & \colhead{} & $  0.523^{+  0.187}_{ -0.219}$ & $  0.276^{+  0.209}_{ -0.193}$ & $  0.341^{+  0.16}_{ -0.202}$ & \colhead{} & $  0.471^{+  0.174}_{ -0.186}$ & $  0.269^{+  0.147}_{ -0.153}$ & $  0.325^{+  0.12}_{ -0.136} $ \\
NGC 5831 &   & $  0.129^{+  0.208}_{ -0.084}$ & $  0.625^{+  0.033}_{ -0.111}$ & \colhead{} & $  0.123^{+  0.225}_{ -0.082}$ & $  0.679^{+  0.036}_{ -0.142}$ & \colhead{} & $  0.200^{+  0.203}_{ -0.146}$ & $  0.494^{+  0.083}_{ -0.106}$ & $  0.563^{+  0.055}_{ -0.089}$ & \colhead{} & $  0.172^{+  0.222}_{ -0.109}$ & $  0.484^{+  0.128}_{ -0.087}$ & $  0.550^{+  0.097}_{ -0.067} $ \\
NGC 5846 &   & $  0.444^{+  0.322}_{ -0.164}$ & $  0.674^{+  0.115}_{ -0.518}$ & \colhead{} & $  0.185^{+  0.437}_{ -0.124}$ & $  0.868^{+  0.054}_{ -0.327}$ & \colhead{} & $  0.587^{+  0.138}_{ -0.179}$ & $  0.526^{+  0.156}_{ -0.216}$ & $  0.586^{+  0.134}_{ -0.175}$ & \colhead{} & $  0.438^{+  0.217}_{ -0.271}$ & $  0.622^{+  0.213}_{ -0.186}$ & $  0.672^{+  0.191}_{ -0.178} $ \\
NGC 5869 &   & $  0.100^{+  0.188}_{ -0.072}$ & $  0.720^{+  0.023}_{ -0.087}$ & \colhead{} & $  0.094^{+  0.206}_{ -0.068}$ & $  0.756^{+  0.032}_{ -0.109}$ & \colhead{} & $  0.134^{+  0.166}_{ -0.092}$ & $  0.583^{+  0.084}_{ -0.068}$ & $  0.644^{+  0.048}_{ -0.048}$ & \colhead{} & $  0.114^{+  0.204}_{ -0.077}$ & $  0.575^{+  0.114}_{ -0.079}$ & $  0.638^{+  0.079}_{ -0.059} $ \\
NGC 6703 &   & $  0.090^{+  0.084}_{ -0.057}$ & $  0.732^{+  0.018}_{ -0.031}$ & \colhead{} & $  0.101^{+  0.097}_{ -0.066}$ & $  0.689^{+  0.028}_{ -0.036}$ & \colhead{} & $  0.115^{+  0.136}_{ -0.077}$ & $  0.578^{+  0.087}_{ -0.057}$ & $  0.650^{+  0.044}_{ -0.031}$ & \colhead{} & $  0.128^{+  0.171}_{ -0.078}$ & $  0.541^{+  0.08}_{ -0.058}$ & $  0.619^{+  0.04}_{ -0.053} $ \\
\enddata
\tablecomments{Fitted values for the four different cases of Table~\ref{tab:chisq}. Parameter $\Upsilon_{\star 0}$ (Equation~(\ref{eq:MLgrad})) refers to the value of $M_\star/L$ for the region where $M_\star/L$ is constant while $\Upsilon_{\star{\rm e}}$ is the average $M_\star/L$ for the projected region within $R_{\rm e}$, which is of course the same as $\Upsilon_{\star 0}$ for the cases of (a) and (b). Parameter $f_{\rm DM}$ refers to the fraction of dark matter within the spherical volume of radius $r=R_{\rm e}$.}
\end{deluxetable*}
\end{longrotatetable}


\begin{thebibliography}{}

\bibitem[Alton et al.(2017)]{Alt17} Alton, P. D., Smith, R. J., Lucey, J. R. 2017, \mnras, 468, 1594

\bibitem[Alton et al.(2018)]{Alt18} Alton, P. D., Smith, R. J., Lucey, J. R. 2018, \mnras, 478, 4464

\bibitem[Balcells \& Quinn(1990)]{BQ90} Balcells, M., Quinn, P. J. 1990, \apj, 361, 381  

\bibitem[Barber et al.(2018)]{Bar18} Barber, C, Crain, R. A., Schaye, J. 2018, \mnras, 479, 5448
  
\bibitem[Barnes \& Hernquist(1996)]{BH96} Barnes, J. E., Hernquist, L. 1996, \apj, 471, 115  

\bibitem[Bernardi et al.(2018)]{Ber18} Bernardi, M., Sheth, R. K., Dominguez-Sanchez, H., et al.\ 2018, MNRAS, 477, 2560
  
\bibitem[Binney \& Mamon(1982)]{BM82} Binney, J, Mamon, G.~A. 1982, \mnras, 200, 361

\bibitem[Binney \& Tremaine(2008)]{BT} Binney, J., Tremaine, S. 2008, Galactic Dynamics, (2nd ed.; Princeton, NJ: Princeton Univ.\ Press)

\bibitem[Bois et al.(2011)]{Bois11} Bois, M., et al.\ 2011, \mnras, 416, 1654 
  
\bibitem[Bundy et al.(2015)]{Bun15} Bundy, K., et al.\ 2015, \apj, 798, 7

\bibitem[Cappellari(2016)]{Cap16} Cappellari, M. 2016, \araa, 54, 597  

\bibitem[Cappellari et al.(2006)]{Cap06} Cappellari, M., Bacon, R., Bureau, M., et al.\ 2006, \mnras, 366, 1126

\bibitem[Cappellari et al.(2007)]{Cap07} Cappellari, M., Emsellem, E., Bacon, R., et al.\ 2007, \mnras, 379, 418

\bibitem[Cappellari et al.(2011)]{Cap11} Cappellari, M., Emsellem, E., Krajnovi\'{c}, D., et al.\ 2011, \mnras, 413,
 813

\bibitem[Cappellari et al.(2013a)]{Cap13} Cappellari, M., Scott, N., Alatalo, K., et al.\ 2013a, \mnras, 432, 1709

\bibitem[Cappellari et al.(2013b)]{Cap13b} Cappellari, M., McDermid, R. M., Alatalo, K., et al.\ 2013b, \mnras, 432, 1862

\bibitem[Chae, Bernardi \& Sheth(2018)]{CBS18a} Chae, K.-H., Bernardi, M., Sheth,  R. K. 2018, \apj, 860, 81 (Paper I)

\bibitem[Chae et al.(2019)]{CBS18b} Chae, K.-H., Bernardi, M., Sheth,  R. K., Gong, I.-T. 2019, \apj, submitted
  
\bibitem[Davis \& McDermid(2017)]{DM17} Davis, T. A., McDermid, R. M. 2017, \mnras, 464, 453

\bibitem[de Zeeuw(1985)]{deZ85} de Zeeuw, T. 1985, \mnras, 216, 273
  
\bibitem[de Zeeuw et al.(2002)]{deZ02} de Zeeuw, P. T., et al.\ 2002, \mnras, 329, 513
  
\bibitem[Einasto(1965)]{Ein} Einasto, J. 1965, TrAlm, 5, 87  
  
\bibitem[Emsellem et al.(2007)]{Ems07} Emsellem, E., Cappellari, M., Krajnovi\'{c}, D., et al.\ 2007, \mnras, 379, 401

\bibitem[Emsellem et al.(2011)]{Ems11} Emsellem, E., Cappellari, M., Krajnovi\'{c}, D., et al.\ 2011, \mnras, 414, 888 
  
\bibitem[Famaey \& Binney(2005)]{FB05} Famaey, B., Binney, J. 2005, \mnras, 363, 603

\bibitem[Gebhardt et al.(2003)]{Geb03} Gebhardt, K., Richstone, D., Tremaine, S., et a..\ 2003, \apj, 583, 92

\bibitem[Genel et al.(2014)]{Gen14} Genel, S., et al.\ 2014, \mnras, 445, 175 
    
\bibitem[Gerhard et al.(2001)]{Ger01} Gerhard, O., Kronawitter, A., Saglia, R. P., Bender, R. 2001, \aj, 121, 1936

\bibitem[Hilz et al.(2012)]{Hilz12} Hilz, M., Naab, T., Ostriker, J. P., Thomas, J., Burkert, A., Jesseit, R.\ 2012, \mnras, 425, 3119   

\bibitem[Janz et al.(2016)]{Jan16} Janz, J., Cappellari, M., Romanowsky, A. J.,  Ciotti, L., Alabi, A., Forbes, D. A. 2016, \mnras, 461, 2367  
  
\bibitem[Jesseit et al.(2005)]{Jess05} Jesseit, R., Naab, T., Burkert, A.\ 2005, \mnras, 360, 1185

\bibitem[Jesseit et al.(2007)]{Jess07} Jesseit, R., Naab, T., Peletier, R. F., Burkert, A.\ 2007, \mnras, 376, 997

\bibitem[Jorgensen et al.(1995)]{Jor95} Jorgensen, I., Franx, M., Kjaergaard, P. 1995, \mnras, 276, 1341  
  
\bibitem[Kent(1987)]{Ken87} Kent, S. M. 1987, \aj, 93, 816

\bibitem[Koopmans et al.(2009)]{Koo09} Koopmans, L. V. E., Bolton, A., Treu, T., et al.\ 2009, \apj, 703, L51

\bibitem[Kormendy(2016)]{Kor16} Kormendy, J. 2008 in Galactic Bulges (Astrophysics and Space Science Library, Vol. 418: Springer International Publishing Switzerland), p.\ 431

\bibitem[Krajnovi\'{c} et al.(2005)]{Kra05} Krajnovi\'{c}, D., Cappellari, M., Emsellem, E., McDermid, R. M., de Zeeuw, P. T. 2005, \mnras, 357, 1113 
  
\bibitem[Krajnovi\'{c} et al.(2008)]{Kra08} Krajnovi\'{c}, D., Bacon, R., Cappellari, M., et al.\ 2008, \mnras, 390, 93 
  
\bibitem[Krajnovi\'{c} et al.(2011)]{Kra11} Krajnovi\'{c}, D., Emsellem, E., Cappellari, M., et al.\ 2011, \mnras, 414, 2923

\bibitem[Krajnovi\'{c} et al.(2013)]{Kra13} Krajnovi\'{c}, D., Alatalo, K., Blitz, L., et al.\ 2013, \mnras, 432,  1768

\bibitem[La Barbera et al.(2016)]{LaB16} La Barbera, F., Vazdekis, A., Ferreras, I., et al., 2016, \mnras, 457, 1468
  
\bibitem[Li et al.(2018)]{Li18} Li. H., Mao, S., Emsellem, E., et al.\ 2018, \mnras, 473, 1489

\bibitem[Mart\'{i}n-Navarro et al.(2015)]{MN15} Mart\'{i}n-Navarro, I., La Barbera, F., Vazdekis, A., Falc\'{o}n-Barroso, J., Ferreras, I. 2015, \mnras, 447, 1033

\bibitem[McGaugh(2008)]{McG08} McGaugh, S. 2008, \apj, 683, 137

\bibitem[Merritt(1985)]{Merr} Merritt, D. 1985, \aj, 90, 1027

\bibitem[Merritt et al.(2006)]{Merr06} Merritt, D., Graham, A. W., Moore, B., Diemand, J., Terzi\'{c}, B. 2006, \aj, 132, 2685 

\bibitem[Milgrom(1983)]{Mil} Milgrom, M. 1983, \apj, 270, 371

\bibitem[Mo, van den Bosch \& White(2010)]{MBW10} Mo, H., van den Bosch, F. C., White, S. 2010, Galaxy Formation and Evolution (Cambridge: Cambridge Univ.\ Press)

\bibitem[Naab et al.(2014)]{Naab14} Naab, T., Oser, L., Emsellem, E., et al.\ 2014, \mnras, 444, 3357  
  
\bibitem[Navarro, Frenk \& White(1997)]{NFW} Navarro, J. F., Frenk, C. S., White, S. D. M. 1997, \apj, 490, 493

\bibitem[Navarro et al.(2010)]{Nav10} Navarro, J.~F., Ludlow, A., Springel, V., et al.\ 2010, \mnras, 402, 21   

\bibitem[Oldham \& Auger(2018)]{Old18} Oldham, L., Auger, M. 2018, \mnras, 474, 4169  

\bibitem[Oser et al.(2010)]{Oser10} Oser, L., Ostriker, J. P., Naab, T., Johansson, P. H., Burkert, A. 2010, \apj, 725, 2312  

\bibitem[Osipkov(1979)]{Osip} Osipkov, L. P. 1979, Pis'ma v Astron. Zhur., 5, 77
  
\bibitem[Richstone \& Tremaine(1988)]{RT88} Richstone, D. O., Tremaine, S. 1988, \apj, 327, 82  

\bibitem[R\"{o}ttgers et al.(2014)]{Rott14} R\"{o}ttgers, B., Naab, T., Oser, L. 2014, \mnras, 445, 1065

\bibitem[Sarzi et al.(2018)]{Sar18} Sarzi, M., Spiniello, C., La Barbera, F., Krajnovi\'{c}, D., van den Bosch, R. 2018, \mnras, 478, 4084

\bibitem[S\'{e}rsic(1968)]{Ser} S\'{e}rsic, J. L. 1968, Atlas de Galaxias Australes (C\'{o}rdoba: Observatorio Astron\'{o}mico)

\bibitem[Sonnenfeld et al.(2018)]{Son18} Sonnenfeld, A., Leauthaud, A.,  Auger, M. W., et al.\ 2018, \mnras, in press

\bibitem[Statler(1987)]{Stat87} Statler, T. S. 1987, \apj, 321, 113
  
\bibitem[Thomas et al.(2007)]{Tho07} Thomas, J., Saglia, R. P., Bender, R., et al.\ 2007, \mnras, 382, 657

\bibitem[Tsatsi et al. (2015)]{Tsat15} Tsatsi, A., Macci\'{o}, A. V., van de Ven, G., Moster, B. P., 2015, \apjl, 802, L3

\bibitem[van der Marel et al.(1998)]{vdM98} van der Marel, R. P., Cretton, N., de Zeeuw, P. T., Rix, H.-W. 1998, \apj, 493, 613
  
\bibitem[van Dokkum et al.(2017)]{vD17} van Dokkum, P., Conroy, C., Villaume, A. , Brodie, J., Romanowsky, A. J. 2017, \apj, 841, 68 

\bibitem[Vogelsberger et al.(2014a)]{Vog14a} Vogelsberger, M., et al.\  2014a, \mnras, 444, 1518

\bibitem[Vogelsberger et al.(2014b)]{Vog14b} Vogelsberger, M., et al.\  2014b, Nature, 509, 177
  
\bibitem[Wu et al.(2014)]{Wu14} Wu, X., Gerhard, O., Naab, T., et al.\ 2014, \mnras, 438, 2701   

\bibitem[Xu et al.(2017)]{Xu17} Xu, D., Springel, V., Sluse, D., et al.\ 2017, \mnras, 469, 1824

\end{thebibliography}
\end{document}